\pdfoutput=1    

\documentclass[pra,aps,twocolumn,superscriptaddress,showpacs,showkeys]{revtex4-1}

\usepackage{amsmath,amsthm,amscd,amsfonts,amssymb}
\usepackage{graphicx}   
\usepackage[usenames,dvipsnames]{xcolor}   
\usepackage{siunitx}                     
\usepackage{hyperref}
\usepackage{caption}             
\usepackage{subcaption}         

\usepackage{calrsfs}      

\renewcommand{\vec}[1]{\mathbf{{#1}}}                         
\newcommand{\pvec}[1]{\mathbf{{#1}}_\parallel }
\newcommand{\vecUnit}[1]{\mathbf{\hat{#1}}}                         
\newcommand{\pvecUnit}[1]{\mathbf{\hat{#1}}_\parallel }
 \newcommand{\imu}{\mathrm{i}}
\newcommand{\dint}[2][]{\!\mathrm{d}^{#1}#2\,}                 
\newcommand{\dfint}[3][]{\! \frac{\mathrm{d}^{#1}#2}{#3}\,}

\newcommand{\etal}{{\it et al.\!}}

\graphicspath{{Figures/},.} 

\begin{document}


\title{Wave scattering from two-dimensional self-affine Dirichlet and  Neumann surfaces and its application to the retrieval of self-affine parameters}

\author{Daniel Strand}
\affiliation{Department of Physics, NTNU --- Norwegian University of Sciences and Technology, NO-7491 Trondheim, Norway}

\author{Torstein Nesse}
\affiliation{Department of Physics, NTNU --- Norwegian University of Sciences and Technology, NO-7491 Trondheim, Norway}

\author{Jacob B. Kryvi}
\affiliation{Department of Physics, NTNU --- Norwegian University of Sciences and Technology, NO-7491 Trondheim, Norway}

\author{Torstein Storflor Hegge}
\affiliation{Department of Physics, NTNU --- Norwegian University of Sciences and Technology, NO-7491 Trondheim, Norway}

\author{Ingve Simonsen}
\affiliation{Department of Physics, NTNU --- Norwegian University of Sciences and Technology, NO-7491 Trondheim, Norway}
\affiliation{Department of Petroleum Engineering, University of Stavanger, NO-4036 Stavanger, Norway}
\affiliation{Surface du Verre et Interfaces, UMR 125 CNRS/Saint-Gobain, F-93303 Aubervilliers, France}

\begin{abstract}
  Wave scattering from two-dimensional self-affine Dirichlet and Neumann surfaces is studied for the purpose of using the intensity scattered from them to obtain the Hurst exponent and topothesy that characterize the self-affine roughness. By the use of the Kirchhoff approximation a closed form mathematical expression for the angular dependence of the mean differential reflection coefficient is derived under the assumption that the surface is illuminated by a plane incident wave. It is shown that this quantity can be expressed in terms of the isotropic, bivariate ($\alpha$-stable) L\'evy distribution of a stability parameter that is two times the Hurst exponent of the underlying surface. Features of the expression for the mean differential reflection coefficient are discussed, and its predictions compare favorably over large regions of parameter space to results obtained from rigorous computer simulations based on equations of scattering theory. It is demonstrated how the Hurst exponent and the topothesy of the self-affine surface can be inferred from scattering data it produces. Finally several possible scattering configurations are discussed that allow for an efficient extraction of these self-affine parameters.
\end{abstract}
\keywords{self-affinity, randomly rough surface, rough surface scattering, inverse scattering problem, surface height autocorrelation function}

\date{\today}
\maketitle


\section{Introduction}

Research on the scattering of waves from rough surfaces dates back as far as to the 1890s when Lord Rayleigh conducted a series of seminal studies on the topic~\cite{Book:Rayleigh1945}. Since then much progress has been made on the experimental and theoretical aspects of the problem~\cite{Book:Bass1979,Book:Beckmann1987,Book:Ogilvy1991,Book:Chew1995,Book:Voronovich1998,Warnick2001,Book:Nieto-Vesperinas2006,Simonsen2004-3}. Today surfaces with well controlled statistical roughness can be manufactured to facilitate the comparison between experimental and theoretical predictions~\cite{ODonriell1987,Navarrete2002}, and the full angular distribution of the scattered intensity can be measured~\cite{Chaikina2000,Chaikina2002} and calculated for both metallic and dielectric surfaces~\cite{Soriano2001,Simonsen2009-1,Simonsen2009-9,Simonsen2010-04,Simonsen2011-05,Simonsen2014-01,Simonsen2013-05}. Initially, the theoretical treatment was concentrated around various perturbation theories~\cite{Jackson1988,Book:Ogilvy1991,Maradudin1993a,Elfouhaily2004,Berginc2007} and single scattering approximations, like the Kirchhoff approximation~\cite{Voronovich2007,Ishimaru1990,Bruce2011,Caron2002,Soriano2002}, methods that are expected to be accurate for weakly rough surfaces and/or surfaces of small slopes. With the advent of the computer, non-perturbative, purely numerical solutions of the scattering problem started to become practically possible from the last half of the 1980s. The first such simulations focused on the scattering from one-dimensional random surfaces~\cite{Warnick2001,Johnson2007,Simonsen2004-3,Maradudin1990,Simonsen1998-4} and only rather recently, due to its numerical complexity, has wave scattering from two-dimensional randomly rough surfaces been tackled by rigorous numerical methods~\cite{Tran1992,Macaskill1993,Tran1993,Warnick2001,Saillard2001,Soriano2002,Simonsen2009-1,Simonsen2009-9,Book:Chew2009,Simonsen2010-04,Simonsen2011-05,Simonsen2014-01,Simonsen2013-05}. 
 
Scattering of waves from ordered or disordered rough surfaces is of interest in various fields of science, engineering, and medicine. For instance, X-rays are used routinely in material science to uncover the underlying crystal structure of materials, and in medicine and dentistry as a diagnostic tool. Inverse scattering techniques are used in geophysical exploration for reservoirs of hydrocarbons and fresh water, ground penetrating radar is used for imaging the subsurface for the purpose of locating archaeological artifacts or mines, and ultrasound imaging is a safe and non-invasive medical technique used to image the inside of the body using sound waves~\cite{Book:Colton2012}. Recently, an inverse scattering technique was developed that uses electromagnetic waves for the purpose of extracting the statistical properties of two-dimensional randomly rough surfaces from the knowledge of the in-plane and co-polarized  scattered intensity distribution~\cite{Simonsen2014-05}. The advantage of using wave based methods for this purpose relative to say scanning probe methods, like for instance atomic force microscopy or contact profilometers, is that it is nondestructive and may cover large surface areas in a short amount of time which is essential when the information that one seeks is statistical in nature. The purpose of this paper is to develop similar capabilities for two-dimensional scale invariant rough surfaces.

Scale invariance is a concept that was pioneered by Mandelbrot~\cite{Book:Mandelbrot1983,Book:Feder1988}, and for surfaces, it takes the form of \textit{self-affinity}. It expresses itself as an invariance under scaling (or dilations) that are \textit{different} in the horizontal plane and in the vertical direction, \textit{i.e.} self-affinity is about invariance under anisotropic scaling. Such transformations are in the language of geometry known as affine-transformations~\cite{Book:Berger1987}. Surfaces that are invariant under affine-transformations are known as self-affine surfaces and they are characterized by the roughness or \textit{Hurst exponent}, and a length scale known as the \textit{topothesy} that controls the amplitude of the self-affine surface in much the same way as the standard deviation does for more ``classic'' rough surfaces~\cite{Book:Feder1988}. In addition to these two parameters, any real self-affine surface will in addition require two length scales that characterize the lower and upper cut-offs in the self-affine scaling regime. 

Self-affine surfaces are abundant in nature but also many industrial and other man-made surfaces display self-affine scaling. Some examples are fractured surfaces in a wide range of materials~\cite{Book:Biswas2015,Mandelbrot1984,Maloy1992,Bouchaud1997,Bouchaud2014}; geological structures covering orders of magnitude in length scales~\cite{Candela2009,Renard2004};
the topography of the sea floor~\cite{Book:Turcotte2007}; surfaces resulting from interface growth and roughening phenomena~\cite{Book:Barabasi1995}; and the surfaces of cold-rolled aluminum sheets used, for instance, in the building and construction industry for making building facades~\cite{Plouraboue1999}, to mention a few.

Numerous methods have been proposed in the literature for measuring the Hurst exponents of self-affine surfaces~\cite{Book:Feder1988,Peng1994,Simonsen1998-1}. The majority of these methods are direct methods in the sense that they require that the surface profile function first is measured, often  over a uniform grid of points in the horizontal plane. Moreover, the uncertainty of the obtained estimate for the Hurst exponent depends partly on the number of points and area covered by the grid over which the surface is measured~\cite{Schmittbuhl1995,Bakke2007}. In addition, the methods used to determine the Hurst exponent have their own biases, and it is generally recommended to compare the predictions obtained by several methods~\cite{Schmittbuhl1995}. Recently it was also studied how the size of the tip of the stylus used in measuring the surface topography introduces artifacts into the measurements and thus the estimation of the Hurst exponent~\cite{Lechenault2010}. Almost without exception, the available methods focus on the determining of the Hurst exponent of the self-affine surface and only rarely, is its topothesy reported. In addition to the Hurst exponent, to simultaneously, or independently, also be able to determine the topothesy of the self-affine surface is of advantage since it will aid in better analyzing measured topography maps in a reliable manner~\cite{Lechenault2010}. 

\medskip
Wave scattering from scale invariant surfaces was first studied by M.V.~Berry in the late 1970s~\cite{Berry1979} in a paper where he coined the term \textit{diffractals} to mean wave diffraction from fractals. In this study the author analyzed the complex nature of the diffractals (the scattered intensity) and showed that the intensity drops off as a power-law of an exponent that depends on the fractal dimension of the surface away from the specular direction. For non-fractal surfaces, the similar drop in intensity typically is of an exponential-like character. Since the publication of this seminal work by Berry, numerous studies have been published on related aspects of the problem, some of which can be found in
Refs.~\cite{Berry1981,Jakeman1982,Allain1986,Sinha1988,Yordanov1994a,Guerin1996,Sanchez-Gil1997,Sanchez-Gil1998,Sheppard1996a,Jaggard1990,Lin1995,Simonsen1999-1,Simonsen1999-3,Guerin2001,Simonsen2002-2,Guerin2003,Sanchez-Gil2007,Zamani2012a,Yanguas-Gil2011,Safdari2015}. The overall majority of these studies are either purely numerical and/or the scattered intensity is not obtained in a closed form expression but rather more typically as an infinite series where the terms depend on the self-affine parameters. This latter situation makes it more challenging to uncover the relevant self-affine parameters of the surface from the measured scattering data. There are, however, a few noticeable exception to this rule for one-dimensional surfaces, that is, surface roughness that is constant along one direction. In the studies reported in  Refs.~\cite{Simonsen1999-1,Simonsen1999-3} it was demonstrated that the electromagnetic scattering from self-affine surfaces can be expressed as a closed form expression in terms of the (univariate) L\'evy distribution, also known as the $\alpha$-stable distribution. By the use of the well-known expansions of this distribution around zero and for large values of its argument, the behavior of the scattered intensity around the specular direction and in its diffuse tails were obtained including the prefactors that depend on the self-affine parameters of the surface. Soon thereafter, this formalism was applied successfully for the inversion of measured optical scattering data obtained from a self-affine aluminum sample with respect to the self-affine parameters of its surface~\cite{Simonsen2002-2}. The self-affine parameters obtained in this way, that is, the Hurst exponent and the topothesy, were found to be consistent with the corresponding values obtained by directly analyzing the surface morphology of the sample measured by atomic force microscopy.
  
The purpose of this work is to extend the formalism developed for rough self-affine profiles in Refs.~\cite{Simonsen1999-1,Simonsen1999-3} so that it can handle the much more practically relevant situation of scattering from isotropic, two-dimensional self-affine surfaces.  It should be mentioned that this work was initially motivated by a question raised by one of our experimental colleagues that wanted to measure the Hurst exponent and topothesy of a soft and relaxing fractured clay sample. For this purpose contact methods are less than ideal, and the time it takes to perform the measurements is critical in order to obtain reliable result.

%
%
\smallskip
The remaining part of this paper is organized in the following way. In Sec.~\ref{Sec:Scattering_Geometry} we present the scattering geometry that we will be concerned with, including an introduction to self-affine surfaces and their scaling properties. Then relevant parts of scattering theory are presented in Sec.~\ref{Sec:Scattering_Theory}. In the following section, Sec.~\ref{Sec:MDRC}, the analytic expression for the mean differential reflection coefficient (scattered intensity) is derived within the Kirchhoff approximation, and the prominent features that can be obtained from this expression is discussed. In Sec.~\ref{Sec:Results} we compare the predictions obtained on the basis of the analytic expression with results obtained by rigorous computer simulations. Moreover, in this section it is also discussed how the self-affine parameters, the Hurst exponent and the topothesy, can be determined from in-plane scattering measurements and what are the geometry best suited for doing so. The conclusion that can be drawn from this study is presented in Sec.~\ref{Sec:Conculusions}.

\section{The Scattering Geometry}
\label{Sec:Scattering_Geometry}

\begin{figure}[tbph] 
  \centering
  \includegraphics[width=0.8 \columnwidth]{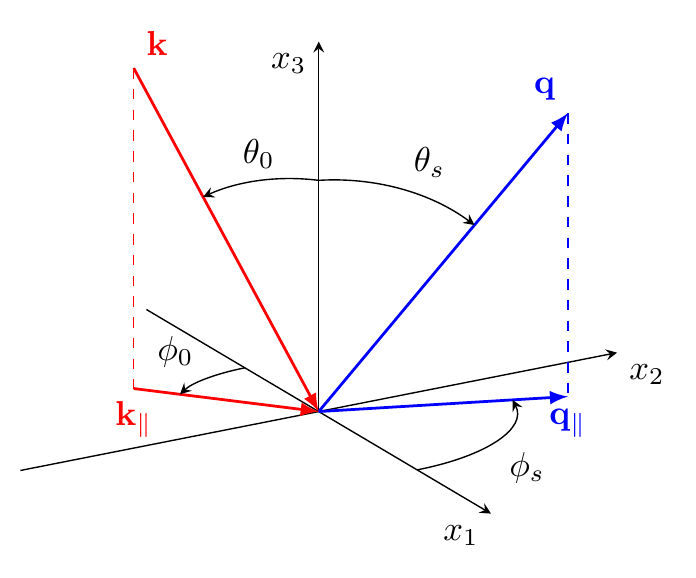}
  \caption{Schematics of the scattering geometry that we consider in this work.}
  \label{fig:scattering_geomatry}
\end{figure}
%

The scattering system that we consider in this work is depicted in Fig.~\ref{fig:scattering_geomatry}.
It consists of a medium where scalar waves can propagate without absorption in the region $x_3>\zeta(\pvec{x})$, and a medium that is impenetrable to scalar waves in the region $x_3<\zeta(\pvec{x})$. Here $\pvec{x}=(x_1,x_2,0)$ represents the position vector in the plane $x_3=0$, and the surface profile function $x_3=\zeta(\pvec{x})$ is assumed to be a single-valued function of $\pvec{x}$ that is differentiable with respect to $x_1$ and $x_2$. It is assumed to constitute a stochastic, isotropic random process that shows \textit{self-affine scaling} and is flat on average, \textit{i.e.}, $\left< \zeta(\pvec{x})\right>=0$ where the angle brackets denote an average over the ensemble of surface realizations.

\medskip
We start the discussion of self-affinity by considering two arbitrarily chosen points, $\pvec{x}$ and $\pvec{x}'$, at which positions the surface profile takes the values $\zeta(\pvec{x})$ and $\zeta(\pvec{x}')$, respectively. These two points are separated by the distance $\Delta \pvec{x}  = \pvec{x}' - \pvec{x}$ in the mean plane, and the corresponding height difference between them is  $\zeta(\pvec{x}') - \zeta(\pvec{x})$ which we in the following denote $\Delta\zeta(\Delta\pvec{x})$.
In a statistical sense, and due to the isotropy of the surface, this height difference will only depend on $\Delta x_\parallel =|\Delta \pvec{x}|$.  If the distance $\Delta \pvec{x}$ is re-scaled to $\Lambda \Delta \pvec{x}$, where $\Lambda$ is a positive constant, then the corresponding height difference, $\Delta\zeta(\Lambda\Delta \pvec{x})$, will be statistically equivalent to $\Lambda^H \Delta\zeta(\Delta x_\parallel)$ with $H>0$ if the surface is self-affine and isotropic. Thus self-affine scaling is defined as  statistical invariance under the transformation (or scaling)~\cite{Book:Feder1988}
\begin{subequations}
  \label{eq:self-affine_scaling}
  \begin{align}
    \Delta \pvec{x}
    &\rightarrow  \Lambda \, \Delta \pvec{x}
    \\
    \Delta \zeta(  \Delta \pvec{x})
    &\rightarrow 
      \Lambda^H \, \Delta \zeta( \Delta \pvec{x}).
\end{align}
\end{subequations}
In writing Eq.~\eqref{eq:self-affine_scaling} we have introduced the so-called Hurst exponent (or roughness exponent), $H$, that can take values in the interval $0<H<1$. This parameter characterizes the self-affinity of the surface $\zeta(\pvec{x})$. It can be shown that if $H>1/2$, the height differences are positively correlated a situation referred to as persistent self-affine surfaces; on the other hand, if  $H<1/2$ the height differences are negatively correlated and one talks about anti-persistent self-affine surfaces~\cite{Book:Feder1988}. Finally, when $H=1/2$ the height differences are uncorrelated  and the process is of the random walk (or Brownian) type.

In what follows, it will be of interest to know the typical slope of the surface. To this end, we start by defining the root-mean-square~(rms) height difference of the surface over a lateral distance $\Delta x_\parallel=|\Delta \pvec{x}|$
\begin{align}
  \label{eq:rms-deviation-def}
  \sigma( \Delta x_\parallel ) =  \left< \left[ \zeta(\pvec{x}+\Delta \pvec{x}) - \zeta(\pvec{x}) \right]^2 \right>^{1/2}_{\pvec{x}}.
\end{align}
Here, $\left< \cdot \right>_{\pvec{x}}$ signifies an average with respect to $\pvec{x}$. From the self-affine scaling relation~\eqref{eq:self-affine_scaling}, it readily follows that $\sigma(\Delta x_\parallel) \simeq \Lambda^{-H}\,\sigma(\Lambda\,\Delta x_\parallel)$ where the symbol $\simeq$ is used to mean ``equivalent in a statistical sense''. Introducing a lateral length scale --- the \textit{topothesy}
 ---  denoted by the symbols $\ell$ and defined so that $\sigma(\ell)\equiv \ell$, one finds that 
\begin{align}
    \label{eq:rms-deviation}
  \sigma(\Delta x_\parallel ) = \ell^{1-H} \Delta x_\parallel^{H}.
\end{align}
The smaller the topothesy, the flatter the surface looks at the macroscopic scale.  With Eq.~\eqref{eq:rms-deviation}, the rms-slope of the surface calculated over a distance $\Delta x_\parallel$, becomes 
\begin{align}
  \label{eq:rms-slope}
  s(\Delta x_\parallel) &= \frac{ \sigma(\Delta x_\parallel) }{\Delta x_\parallel} 
                        = \left( \frac{\ell}{\Delta x_\parallel} \right)^{1-H}. 
\end{align}
Equation~\eqref{eq:rms-slope} predicts that the rms-slope is less than one for $\Delta x_\parallel > \ell$; it is larger than one for $\Delta x_\parallel < \ell$; while at $\Delta x_\parallel = \ell$ one has $ s(\ell)=1$. This result allows the geometrical interpretation of the topothesy as the length scale in the mean plane over which the surface has an average slope of one (or \ang{45}). At least in a box-counting sense, the self-affine surface is fractal \emph{only} for (lateral) length scales  $\Delta x_\parallel < \ell$~\cite{Book:Feder1988} but it is self-affine at any length scales~\footnote{This is true only as long as no cut-off length-scales have been introduced that can restrict the region of self-affine scaling.}. Therefore, the physical significance of the topothesy, $\ell$, is to distinguish the fractal region from the non-fractal region of the self-affine surface.

The self-affine scaling in Eq.~\eqref{eq:self-affine_scaling} is often written in the more compact form $\zeta(  \Delta \pvec{x}) \simeq \Lambda^{-H} \zeta(\Lambda\Delta \pvec{x})$, where we recall that $\simeq$ means ``statistically equivalent''. For instance, statistical equivalence means that the probability density function, $p(\Delta \zeta; \Delta x_\parallel)$, of finding a height difference $\Delta \zeta$ over a lateral distance $\Delta x_\parallel=|\Delta\pvec{x}|$, has to satisfy the relation
\begin{align}
  \label{eq:pdf-scaling}
  p(\Delta \zeta; \Delta x_\parallel)  = \Lambda^{H}   p(\Lambda^H \Delta \zeta; \Lambda \Delta x_\parallel),
\end{align}
which is a consequence of the transformation of random variables~\cite{Book:Ross2002}.  By assuming that this pdf has a Gaussian form, one finds that it is given as
\begin{align}
  \label{eq:pdf-height-diff}
  p(\Delta \zeta; \Delta x_\parallel) 
  &=
  \frac{1}{\sqrt{2\pi} \ell^{1-H} \Delta x_\parallel^H}
  \exp \left[ -\frac{1}{2} \left( \frac{ \Delta \zeta }{ \ell^{1-H} \Delta x_\parallel^H } \right)^2 \right],
\end{align}
where the expression for $\sigma(\Delta x_\parallel )$ given by Eq.~\eqref{eq:rms-deviation} can be recognized in the denominators of both the exponent and the prefactor of the exponential function that appear in this expression. It is straightforward to show that the form for $p(\Delta \zeta; \Delta x_\parallel)$  in Eq.~\eqref{eq:pdf-height-diff} satisfies Eq.~\eqref{eq:pdf-scaling}.

\medskip
It remains to mention that for any physical system, self-affine scaling cannot be expected to hold for all length scales. Instead, there has to be a limitation in the range of scales over which the self-affine scaling exists. To this end, one  introduces a lower and an upper length scale cut-off, denoted $\xi_-$ and $\xi_+$, respectively, outside which such scaling does not hold. The topothesy $\ell$ associated with the surface may be in, or outside, the interval $[\xi_-,\xi_+]$. In the latter case we for length scales $\Delta x_\parallel\in[\xi_-,\xi_+]$, deal with a non-fractal self-affine surface when  $\ell<\xi_-$, and a self-affine fractal surface if $\ell> \xi_+$. When $\ell\in[\xi_-,\xi_+]$ the fractal nature of the surface is only present for $\Delta x_\parallel<\ell$. In the following we will assume that $\ell<\xi_-$ since this is the situation for the surfaces that we will be concerned. For such a  discretized self-affine surface covering a square region of the $x_3=0$ plane of area  $L\times L$,  where  $L$ is the length of one of its edges, we note that $\xi_+= L$ and $\xi_-$ is limited downward by the discretization interval that we assumed to be larger than the topothesy.

\section{Scattering theory}
\label{Sec:Scattering_Theory}
The self-affine surface $x_3=\zeta(\pvec{x})$ is illuminated from above by a time-harmonic plane incident scalar wave of angular frequency $\omega$. In the region $x_3>\zeta(\pvec{x})$, the total field is the sum of an incident and a scattered field
\begin{subequations}
  \label{eq:total-field}
\begin{align}
  \psi(\vec{x},t) &= 
  \left[  
    \psi(\vec{x}|\omega)_{\textsl{inc}} +  \psi(\vec{x}|\omega)_{\textsl{sc}} 
  \right]                       
  \exp(-\imu\omega t),
\end{align}
where the incident field, characterized by the lateral wave vector $\pvec{k}$ and wavenumber $k_\parallel=|\pvec{k}|$,  has the form
\begin{align}
  \label{eq:incident-field}
  \psi(\vec{x}|\omega)_{\textsl{inc}} &=
  \exp\left[ \imu \pvec{k}\cdot \pvec{x} - \imu \alpha_0(k_\parallel)x_3\right],
\end{align}
and the scattered field is given by
\begin{align}
  \label{eq:scattered-field}
  \psi(\vec{x}|\omega)_{\textsl{sc}} &=
  \int \dfint[2]{q_\parallel}{(2\pi)^2} 
     R(\pvec{q}| \pvec{k})   
     \exp\left[ \imu \pvec{q}\cdot \pvec{x} + \imu \alpha_0(q_\parallel)x_3\right].
\end{align}
\end{subequations}
In writing Eq.~\eqref{eq:total-field}, we have introduced the \textit{scattering amplitudes} $R(\pvec{q}| \pvec{k})$ from incident lateral wave vector $\pvec{k}$  into scattered lateral wave vectors $\pvec{q}$, and defined
\begin{align}
  \label{eq:alpha0}
  \alpha_0(q_\parallel) &=
  \begin{cases}
    \sqrt{ \frac{\omega^2}{c^2} -q_\parallel^2 },       &   q_\parallel \leq \omega/c\\
    \imu \sqrt{ q_\parallel^2 -\frac{\omega^2}{c^2} }, &   q_\parallel  > \omega/c
  \end{cases},
\end{align}
where $c$ denotes the velocity of propagation of the scalar wave (in the region $x_3>\zeta(\pvec{x})$). The quantity $\alpha_0(q_\parallel)$ represents the 3rd component of the wave vector $\vec{q}$ when the length of its parallel component is $q_\parallel$. The function $\alpha_0(q_\parallel)$ is defined in such a way that the dispersion relation is satisfied for any value of $q_\parallel$. Moreover, it should be pointed out that the form of the scattered field~\eqref{eq:scattered-field}, has also been subjected to an outgoing radiation condition at infinity (Sommerfeld radiation conditions~\cite{Book:Sommerfeld1949,Schot1992}).

The scattering amplitudes, $R(\pvec{q}|\pvec{k})$, that appear in Eq.~\eqref{eq:scattered-field} are important quantities for the following discussion since they are directly related to physical observables. Our prime quantity of interest of this kind is the \textit{mean differential reflection coefficient}~(mean DRC), defined as~\cite[Sec.~3.1]{Book:Maradudin2008}
\begin{align}
  \label{eq:MDRC}
  \left< \frac{\partial R(\pvec{q}|\pvec{k}) }{ \partial \Omega_s} \right>   
     &=
     \frac{1}{S} \left( \frac{\omega}{2\pi c}\right)^2 \frac{\cos^2\theta_s}{\cos\theta_0}
     \left< \left| R(\pvec{q}|\pvec{k}) \right|^2 \right>,
\end{align}
where $S$ is the area of the mean plane covered by the rough surface, and the  lateral wave vectors, $\pvec{k}$ and $\pvec{q}$, are expressed in terms of the angles of incidence 
\begin{subequations}
\label{eq:kq-parallel}
\begin{align}
  \label{eq:k-parallel}
  \pvec{k} &= \frac{\omega}{c} \sin\theta_0 (\cos\phi_0, \sin\phi_0, 0 ),
\end{align}
and the angles of scattering (when $|\pvec{q}|<\omega/c$)  
\begin{align}
\label{eq:q-parallel}
  \pvec{q} &= \frac{\omega}{c} \sin\theta_s (\cos\phi_s, \sin\phi_s, 0 ).
\end{align}
\end{subequations}
Moreover, in the radiative region, one can infer from Eqs.~\eqref{eq:alpha0} and ~\eqref{eq:kq-parallel} that 
\begin{subequations}
  \label{eq:alpha0-radiative}
\begin{align}
  \label{eq:alpha0-radiative-A}
  \alpha_0(k_\parallel) &= \frac{\omega}{c} \cos\theta_0, \qquad k_\parallel < \omega/c
  \\
  \label{eq:alpha0-radiative-B}
  \alpha_0(q_\parallel) &= \frac{\omega}{c} \cos\theta_s, \qquad q_\parallel < \omega/c.
\end{align}
\end{subequations}

The DRC is defined such that  $\left(\partial R(\pvec{q}|\pvec{k})/ \partial\Omega_s\right) \dint\Omega_s$ is the fraction of the total time-averaged energy flux in an incident field, of lateral wave vector $\pvec{k}$, that is scattered into fields, of lateral wave vector $\pvec{q}$, within a solid angle $\dint\Omega_s$ about the scattering direction defined by the polar and azimuthal scattering angles $(\theta_s,\phi_s)$. Since we are dealing with randomly rough surfaces, it is the average of the DRC over an ensemble of surface realizations, denoted $\left< \cdot \right>$, that we are interested in, and this is what leads to the expression~\eqref{eq:MDRC}.

The scattering amplitudes are determined by imposing proper boundary conditions~(BCs) on the rough surface $x_3=\zeta(\pvec{x})$. For an impenetrable substrate, two boundary conditions are of particular interest. The first, is the \textit{Dirichlet boundary condition} (or first-type BC), that is defined by requiring the total field on the surface to vanish; 
\begin{subequations}
  \label{eq:BoundaryConditions}
\begin{align}
  \psi(\vec{x}|\omega)\big|_{x_3 = \zeta(\pvec{x}) }  &= 0.  
\end{align}
The second, is the \textit{Neumann boundary condition} (or second-type BC) which states that the normal derivative of the total field at the surface  should vanish
\begin{align}
  \partial_n\psi(\vec{x}|\omega)
  \big|_{x_3 = \zeta(\pvec{x}) }   &=0,
\end{align}
\end{subequations}
where $\partial_n= \vecUnit{n}(\pvec{x}) \cdot \boldsymbol{\nabla}$ denotes the normal derivative of the surface at $\pvec{x}$ and the unit normal vector at this lateral position is given by
\begin{align}
  \vecUnit{n}(\pvec{x})
  &=
    \frac{
    - \vecUnit{x}_1 \, \partial_1\zeta(\pvec{x})
    - \vecUnit{x}_2 \, \partial_2\zeta(\pvec{x})
    + \vecUnit{x}_3
    }{
    \sqrt{ 1+[\boldsymbol{\nabla} \zeta(\pvec{x}) ]^2 }
    }
\end{align}
where $\partial_i =\partial/\partial x_i$ with $i=1,2$, and a caret over a vector indicates that it is a unit vector.

The scattering problem defined by Eqs.~\eqref{eq:total-field} and \eqref{eq:BoundaryConditions}, can be solved either rigorously, as done in Refs.~\cite{Tran1992,Tran1993,HardWallPaper}, or by adapting various approximations~\cite{Voronovich2007,Book:Maradudin2008}. For instance, as an example of the latter case, within the Kirchhoff approximation the scattering amplitude is given as~\cite{Book:Maradudin2008} 
\begin{widetext}
\begin{equation}
  \label{eq:Kirchhoff}
  \begin{split}
  R(\pvec{q}| \pvec{k}) =
  &\mp 
  \frac{
    \left(\omega/c\right)^2 + \alpha_0(q_\parallel) \alpha_0(k_\parallel) - \pvec{q} \cdot \pvec{k}
    }{
      \alpha_0(q_\parallel) \left[ \alpha_0(q_\parallel) + \alpha_0(k_\parallel)  \right]
  }\\
  & \quad \times \int \dint[2]{x_\parallel} 
    \exp \left\{
      -\imu \left(\pvec{q}-\pvec{k} \right) \cdot \pvec{x}
      -\imu \left[ \alpha_0(q_\parallel) + \alpha_0(k_\parallel) \right] \zeta(\pvec{x})
    \right\},
  \end{split}
\end{equation}
where the upper and lower signs correspond to the Dirichlet and Neumann boundary conditions, respectively. Thus, within the Kirchhoff approximation, the scattering from Dirichlet and Neumann surfaces should give rise to the same mean DRC since this quantity, like any intensity, depends on the absolute square of the scattering amplitude. However, in general it is not true that the Dirichlet and Neumann surfaces scatter scalar waves in the same fashion. To realize this it suffices to note that rough Neumann surfaces can support surface waves~\cite{Zierau2015}, while Dirichlet surfaces cannot.

\section{The mean DRC within the Kirchhoff approximation}
\label{Sec:MDRC} 
If the expression for the scattering amplitude within the Kirchhoff approximation, Eq.~\eqref{eq:Kirchhoff},  is substituted into the expression for the mean DRC, Eq.~\eqref{eq:MDRC}, one obtains
\begin{subequations}
  \label{eq:MDRC-Kirchoff}
\begin{align}
  \label{eq:MDRC-Kirchoff-A}
  \left< \frac{\partial R(\pvec{q}|\pvec{k}) }{ \partial \Omega_s} \right>   
     &=
     \frac{\omega/c}{\alpha_0(k_\parallel)}  
  \frac{ 
    \left[ \left(\omega/c\right)^2 + \alpha_0(q_\parallel) \alpha_0(k_\parallel) - \pvec{q} \cdot \pvec{k} \right]^2
    }{
      \left[ \alpha_0(q_\parallel) + \alpha_0(k_\parallel)  \right]^2
      }
      L( \pvec{q} | \pvec{k} ) ,
\end{align}
where
\begin{align}
  \label{eq:MDRC-Kirchoff-B}
  L( \pvec{q} | \pvec{k} ) 
  &= 
  \frac{1}{ (2\pi)^2 S}
  \int \dint[2]{x_\parallel}
  \int \dint[2]{x_\parallel'}
  \exp \left[ -\imu\left(\pvec{q}-\pvec{k}\right) \cdot (\pvec{x} - \pvec{x}') \right]
     \left<
       \exp \left\{  -\imu \left[ \alpha_0(q_\parallel) + \alpha_0(k_\parallel)   \right] \left[ \zeta(\pvec{x}) - \zeta(\pvec{x}') \right]   \right\}
     \right>,
\end{align}
\end{subequations}
and the results in Eq.~\eqref{eq:alpha0-radiative} have been used.  The average over the surface roughness that appears in Eq.~\eqref{eq:MDRC-Kirchoff-B} is expressed in terms of $\Delta\zeta(\Delta \pvec{x})=\zeta(\pvec{x}) - \zeta(\pvec{x}')$. For a self-affine surface the pdf of these height differences is given by Eq.~\eqref{eq:pdf-height-diff} and a direct calculation that involves the evaluation of a Gaussian integral results in
\begin{align}
\begin{aligned}
  \label{eq:average}
   \left<
       \exp \left\{  -\imu \left[ \alpha_0(q_\parallel) + \alpha_0(k_\parallel)   \right] \left[ \zeta(\pvec{x}) - \zeta(\pvec{x}') \right]   \right\}
     \right> 
   &= 
      \exp 
      \left[
        - \frac{1}{2} \left[ \alpha_0(q_\parallel) + \alpha_0(k_\parallel)   \right]^2 
         \ell^{2-2H} \left| \pvec{x}-\pvec{x}'\right|^{2H}
      \right].
\end{aligned}
\end{align}
As expected, this result shows that the average over an ensemble of realizations of the self-affine surface introduces a dependence  on the topothesy $\ell$ and the Hurst exponent $H$ that characterize the amplitude and correlation, respectively,  of the self-affine surface $x_3=\zeta(\pvec{x})$. On combining Eqs.~\eqref{eq:MDRC-Kirchoff-B} and \eqref{eq:average} and making the change of variable $\pvec{u}=\pvec{x}-\pvec{x}'$ in the first integral of Eq.~\eqref{eq:MDRC-Kirchoff-B}, one obtains
\begin{align}
  \label{eq:L-function}
   L( \pvec{q} | \pvec{k} )
  &=
  \frac{1}{(2\pi)^2}  \int \dint[2]{u_\parallel}    
  \exp \left[ -\imu\left(\pvec{q}-\pvec{k}\right) \cdot \pvec{u} \right]
      \exp 
      \left[
        - \frac{1}{2} \left[ \alpha_0(q_\parallel) + \alpha_0(k_\parallel)   \right]^2 
         \ell^{2-2H} \left|  \pvec{u} \right|^{2H}
      \right],
\end{align}
where it has been used that the second integral in Eq.~\eqref{eq:MDRC-Kirchoff-B} simply evaluates to $S$. 

The function $L( \pvec{q} | \pvec{k} )$ is, in fact, a probability distribution function. It is related to the isotropic \textit{bivariate L\'evy distribution} of stability parameter (or index) $\alpha$ that is centered at zero, and it is defined by~\cite{Press1972,Press1972a,Book:Samoradnitsky1994,Book:Gnedenko1954}
\begin{align}
  \label{eq:Levy}
  {\cal L}_\alpha ( \pvec{Q}; \gamma )
  &=
  \frac{1}{(2 \pi)^2}
  \int \dint[2]{v_\parallel}
  \exp\left( -\imu \pvec{Q} \cdot \pvec{v} \right) 
  \exp \left(-\gamma\left| \pvec{v} \right|^\alpha \right),
  %
  %
\end{align}
with $0<\alpha\leq 2$ and $\gamma>0$. The parameter $\gamma$ is called the \textit{scale} parameter, and it controls the width of the distribution. It should be noted, that in the mathematics and statistics literature, the probability distribution function ${\cal L}_\alpha ( \pvec{Q} ; \gamma )$
is more commonly referred to as the isotropic bivariate $\alpha$-stable distribution.  Only for two values of the stability parameter $\alpha$ can the L\'evy distribution~\eqref{eq:Levy} be expresses in closed form by elementary functions. The first of those values are $\alpha=2$, in which case the resulting distribution is the isotropic bivariate Gaussian distribution of standard deviation $\sqrt{2\gamma}$. The second value is $\alpha=1$, which corresponds to the isotropic bivariate Cauchy-Lorentz distribution (of scale parameter $\gamma$). For all other values of $\alpha\in(0,2]$, the probability distribution function~\eqref{eq:Levy} has to be evaluated numerically, in which case Eq.~\eqref{eq:Levy-Hankel} is useful.  Appendix~\ref{App:Levy} gives  additional details and properties of the L\'evy distribution that will be useful for this work.
It should be noted that only when $\alpha=2$ is the standard deviation of the L\'evy distribution finite; in all other cases $0<\alpha<2$ it is infinite. 
%
%

After combining Eqs.~\eqref{eq:L-function} and \eqref{eq:Levy}, and substituting the resulting expression into Eq.~\eqref{eq:MDRC-Kirchoff}, we find that the mean differential reflection coefficient within the Kirchhoff approximation can be written in terms of the L\'evy distribution of stability parameter, $2H$, as
\begin{subequations}
  \label{eq:MDRC-Kirchoff-Final}
\begin{align}
  \label{eq:MDRC-Kirchoff-Final-A}
  \left< \frac{\partial R(\pvec{q}|\pvec{k}) }{ \partial \Omega_s} \right>   
     &=
  \frac{ 
       \left(\omega/c\right)
      \left[ \left(\omega/c\right)^2 + \alpha_0(q_\parallel) \alpha_0(k_\parallel) - \pvec{q} \cdot \pvec{k} \right]^2
    }{
      \alpha_0(k_\parallel)
      \left[ \alpha_0(q_\parallel) + \alpha_0(k_\parallel)  \right]^2
      }
      {\cal L}_{2H} 
      \left(  
           \pvec{q}-\pvec{k} 
          ;
          \frac{1}{2}
          \left[ \alpha_0(q_\parallel) + \alpha_0(k_\parallel)   \right]^2
          \ell^{2-2H} 
      \right).
\end{align}
Here the quantities $\pvec{k}$, $\alpha_0(k_\parallel)$, $\pvec{q}$, and $\alpha_0(q_\parallel)$ should be understood in terms of the angles of incidence $(\theta_0,\phi_0)$ and the angles of scattering $(\theta_s,\phi_s)$ as given by Eqs.~\eqref{eq:kq-parallel} and \eqref{eq:alpha0-radiative}.

For the later discussion it will be useful to express the L\'evy distribution that appears in Eq.~\eqref{eq:MDRC-Kirchoff-Final-A} in terms of a scaling parameter that is \textit{independent} of both the Hurst exponent and the topothesy that characterize the self-affinity of the rough surface. This is done with the aid of Eq.~\eqref{eq:Levy-scaling} and it is found that the mean DRC alternatively can be expressed as 
\begin{align}
  \label{eq:MDRC-Kirchoff-Final-B}
  \left< \frac{\partial R(\pvec{q}|\pvec{k}) }{ \partial \Omega_s} \right>   
     &=
  \frac{
     \left(\omega/c\right)
    \left[ \left(\omega/c\right)^2 + \alpha_0(q_\parallel) \alpha_0(k_\parallel) - \pvec{q} \cdot \pvec{k} \right]^2
    }{
      \alpha_0(k_\parallel) \left[ \alpha_0(q_\parallel) + \alpha_0(k_\parallel)  \right]^{(2+2H)/H} \ell^{(2-2H)/H}
      }
      {\cal L}_{2H} 
      \left(  
        \frac{ 
          \pvec{q}-\pvec{k} 
        }{ 
          \left[ \alpha_0(q_\parallel) + \alpha_0(k_\parallel)   \right]^{1/H}    
          \ell^{(1-H)/H}     
        }
        ; 
        \frac{1}{2}
      \right).
\end{align}
\end{subequations}
The advantage of this form, over the form in Eq.~\eqref{eq:MDRC-Kirchoff-Final-A}, is that for any self-affine surface, the scale parameter of the L\'evy distribution that appears in Eq.~\eqref{eq:MDRC-Kirchoff-Final-B} is constant and therefore independent of the parameters of the self-affine surface. Moreover, with $\gamma=1/2$ the L\'evy distribution ${\cal L}_2(x;\gamma)$ equals the \textit{standard} Gaussian distribution of zero mean and standard deviation one.

Equation~\eqref{eq:MDRC-Kirchoff-Final} represents a generalization of the results reported previously for the angular distribution for the intensity  scattered from a one-dimensional self-affine randomly rough  surface~\cite{Simonsen1999-1,Simonsen1999-3,Simonsen2002-2}. It is worth noticing that for the one-dimensional case it was also found that within the Kirchhoff approximation the mean differential reflection coefficient can be expressed in terms of a symmetric (univariate) L\'evy distribution of stability parameter $2H$.

\medskip
To aid the subsequent discussion, it will be useful to have available simplified expressions for the intensity distribution~\eqref{eq:MDRC-Kirchoff-Final} around the specular direction $\pvec{q}=\pvec{k}$ and in the diffuse tails of the distribution far away from this direction. The former expression is obtained by first introducing $\pvec{q}=\pvec{k}+\pvec{Q}$ into Eq.~\eqref{eq:MDRC-Kirchoff-Final}, where $\pvec{Q}$ is the lateral wavevector transfer,  and then using the small argument expansion of the L\'evy distribution~\eqref{app:eq:small-argument-expansion} to expand the resulting expression to order $Q_\parallel^2$. In this way, a lengthy but in principle straightforward calculation results in
\begin{align}
  \left< \frac{\partial R( \pvec{k}+\pvec{Q}|\pvec{k})}{\partial \Omega_s}\right>
  &\approx
    \left< \frac{\partial R( \pvec{k}|\pvec{k})}{\partial \Omega_s} \right>
    \left[
    1
    +
    \frac{1-H}{H}
    \frac{ \pvec{k}\cdot\pvec{Q} }{\alpha_0^2(k_\parallel)}
    +
    \frac{6H^2+9H+4}{H^2}
    \left( \frac{\pvec{k}\cdot\pvec{Q} }{\alpha_0^2(k_\parallel)} \right)^2
    \right.
    \nonumber
  \\
  & \qquad \qquad \qquad \qquad \;
    \left.
    -
    \frac{1}{2H}
    \frac{ Q_\parallel^2 }{\alpha_0^2(k_\parallel)}
    \left\{
    1
    +
    H \frac{ \Gamma\left(2/H\right) }{ \Gamma\left(1/H\right) }
    \frac{
    2^{-(1+H)/H}
    }{
    \left[ \alpha_0(k_\parallel) \ell \right]^{(2-2H)/H}
    }   
    \right\}
    \right];
    \qquad
    \frac{Q_\parallel \ell }{ \left[ \alpha_0(k_\parallel)\ell \right]^{1/H} } \ll 1.
    \label{eq:MDRC-specular-expansion-A}
\end{align}
where the mean DRC \textit{at} the specular direction that appears in Eq.~\eqref{eq:MDRC-specular-expansion-A} is defined as
\begin{align}
  \left< \frac{\partial R( \pvec{k}|\pvec{k})}{\partial \Omega_s}\right>
  &=
    2^{-(1+2H)/H}
    \frac{
    \Gamma(1+1/H) 
    }{
    \pi
    }
    \frac{
    \frac{\omega}{c}\ell 
    }{
    \left[ \alpha_0( k_\parallel ) \ell \right]^{(2-H)/H}
    },
    \label{eq:MDRC-specular-direction}
\end{align}
obtained after using the identity $\Gamma(1/H)/H =\Gamma(1+1/H)$.
%
In the following our main concern will be surface roughness for which $\ell/\lambda \ll 1$. If we disregard the possibility of gracing incidence, which is of less practical importance, the condition  $\ell/\lambda \ll 1$ is equivalent to $\alpha_0(k_\parallel)\ell \ll 1$, in which case the dominating $Q_\parallel^2$ term of Eq.~\eqref{eq:MDRC-specular-expansion-A} is the last term in the curly brackets, so that
\begin{align}
  \left< \frac{\partial R( \pvec{k}+\pvec{Q}|\pvec{k})}{\partial \Omega_s}\right>
  &\approx
    \left< \frac{\partial R( \pvec{k}|\pvec{k})}{\partial \Omega_s} \right>
    \left[
    1
    +
    \frac{1-H}{H}
    \frac{ \pvec{k}\cdot\pvec{Q} }{\alpha_0^2(k_\parallel)}
    -
    \frac{\Gamma\left(1/2+1/H\right)}{ 2^{3-1/H} \sqrt{\pi} }
    \frac{
    Q_\parallel^2
    }{
    \alpha_0^2(k_\parallel)
    \left[ \alpha_0(k_\parallel) \ell \right]^{(2-2H)/H}
    }
    \right]
    .
    \label{eq:MDRC-specular-expansion}
\end{align}
\end{widetext}
%
In writing this expression we have used the identity $\Gamma(2/H)/\Gamma(1/H) = 2^{(2-H)/H} \Gamma(1/2+1/H)/\sqrt{\pi}$ known as the duplication formula~\cite[Ch.~5]{NIST:DLMF}. Expression~\eqref{eq:MDRC-specular-expansion} can be used to obtain an estimate for the full width at half maximum~(FWHM) value of the peak, and one finds
\begin{align}
  W(k_\parallel, H, \ell)
  &=
    \sqrt{
    \frac{ 2^{4-1/H} \sqrt{\pi} }{\Gamma\left(1/2+1/H\right)}
    }
    \alpha_0(k_\parallel)
    \left[ \alpha_0(k_\parallel) \ell \right]^{(1-H)/H}
    %
    .
    \label{eq:peak-width}
\end{align}
It should be noted that the nontrivial dependencies on the self-affine parameters that are present in the expressions in Eqs.~\eqref{eq:MDRC-specular-direction} and \eqref{eq:peak-width} can in fact be deduced from simple scaling arguments.
A detailed explanation of how this can be done for one-dimensional self-affine surfaces has already been presented in Ref.~\onlinecite{Simonsen1999-3}, and since these arguments also are valid for the scattering from two-dimensional surfaces, we will here only present the main arguments. The loss of phase coherence (or ``de-phasing'') of the incident beam when it is scattered by a rough surface is due to the competition between two different effects. From the arguments of the two exponential functions that are present in  Eq.~\eqref{eq:MDRC-Kirchoff-B} one finds them to be (\textit{i}) the scattering away from the specular direction, $\pvec{Q}\cdot\Delta\pvec{x}$; and
(\textit{ii}) the scattering from different heights at the
surface, ${Q}_3\Delta\zeta(\Delta \pvec{x})$. Here a wave vector transfer has been defined as  $\vec{Q}=\vec{q}-\vec{k}$ with the incident and scattered wave vectors given as $\vec{k}=\pvec{k}-\vecUnit{x}_3\alpha(k_\parallel)$  and
$\vec{q}=\pvec{q}+\vecUnit{x}_3\alpha(q_\parallel)$, respectively. The two competing effects can be characterized by the two length scales $\chi_\parallel$ and $\chi_3$ that are defined, respectively, via $Q_\parallel \chi_\parallel =2\pi$ and $Q_3 \Delta\sigma(\chi_3)=2\pi$. The transition between the specular regime where $\chi_3/\chi_\parallel\ll 1$, and the diffuse regime for which $\chi_3/\chi_\parallel\gg 1$, takes place when  $\chi_3/\chi_\parallel = 1$. The full width $W$ of the peak in the specular direction is determined from the condition $\chi_3/\chi_\parallel = 1$ and a direct calculation that uses Eq.~\eqref{eq:rms-deviation}, leads to
\begin{align}
  W \sim 2 Q_\parallel
    \propto \alpha_0(k_\parallel ) \left[ \alpha_0(k_\parallel)\ell \right]^{(1-H)/H},
\end{align}
which has the same scaling as Eq.~\eqref{eq:peak-width}.
The amplitude of the mean DRC in the specular direction follows from the energy conservation condition $\int \dint\Omega_s \left< \partial R(\pvec{q}|\pvec{k}/\partial\Omega_s\right>=1$. If it is assumed that most of the intensity scattered by the rough surface ends up inside the region $|\pvec{q}-\pvec{k}|<W/2$, then after using the relation $\textrm{d}^2q_\parallel = (\omega/c)\alpha_0(q_\parallel) \textrm{d}{\Omega_s}$ one finds
\begin{align}
  \left< \frac{\partial R( \pvec{k} |\pvec{k})}{\partial \Omega_s}\right>
  &\sim
    \frac{
    \frac{\omega}{c} \alpha_0( k_\parallel)
    }{
    \pi \left(\frac{W}{2}\right)^2
    }
    \propto
    \frac{
    \frac{\omega}{c} \ell
    }{
    \left[ \alpha_0( k_\parallel) \ell \right]^{(2-H)/H}
    },
\end{align}
which has the same scaling as Eq.~\eqref{eq:MDRC-specular-direction}.

\smallskip
The behavior of the diffuse tails of the scattered intensity distribution~\eqref{eq:MDRC-Kirchoff-Final} is obtained from the large argument asymptotic expansion of the L\'evy distribution, Eq.~\eqref{eq:Levy-large-agument-expansion}, with the result that
\begin{subequations}
  \label{eq:MDRC-diffuse-expansion}
\begin{align}
  \left< \frac{\partial R( \pvec{q} | \pvec{k}) }{ \partial \Omega_s } \right>
  &\simeq
    m(\pvec{q}|\pvec{k})
    \left[ \frac{\Gamma(1+H)}{\pi}\right]^2
    \frac{\sin\left(\pi H\right)}{ \left|\pvec{q}-\pvec{k} \right|^{2+2H}}
    ;
    \nonumber
    \\ & 
    \qquad \qquad \quad  
    \frac{
    \left|\pvec{q}-\pvec{k} \right|
    }{
    \frac{\omega}{c} \left[ \frac{\omega}{c} \ell\right]^{(1-H)/H}
    }
   \gg 1,
    \label{eq:MDRC-diffuse-expansion-a}
\end{align}
where a geometric factor has been defined by
\begin{align}
  m(\pvec{q}|\pvec{k})
  &=
    \frac{\omega/c}{ \alpha_0(k_\parallel)}
    \left[
    \left(\omega/c \right)^2
    +
    \alpha_0(q_\parallel)\alpha_0(k_\parallel)
    -
    \pvec{q}\cdot\pvec{k}
    \right]^2.
        \label{eq:MDRC-diffuse-expansion-b}
\end{align}
Also the behavior  $\left< \partial R( \pvec{q} | \pvec{k}) / \partial \Omega_s  \right>/m(\pvec{q}|\pvec{k})\sim |\pvec{q}-\pvec{k}|^{-2-2H}$ can be obtained from scaling analysis. This can be done by noting that the power spectrum of a two-dimensional self-affine surface satisfies $g(\pvec{k})\sim k_\parallel^{-2-2H}$~\cite{Simonsen2004-3} and use it to repeat the arguments of Ref.~\cite[Sec.~4]{Simonsen1999-3}.
\end{subequations}
%

%
%
\begin{figure}[tb!]
  \centering
   \includegraphics[width=0.45\textwidth]{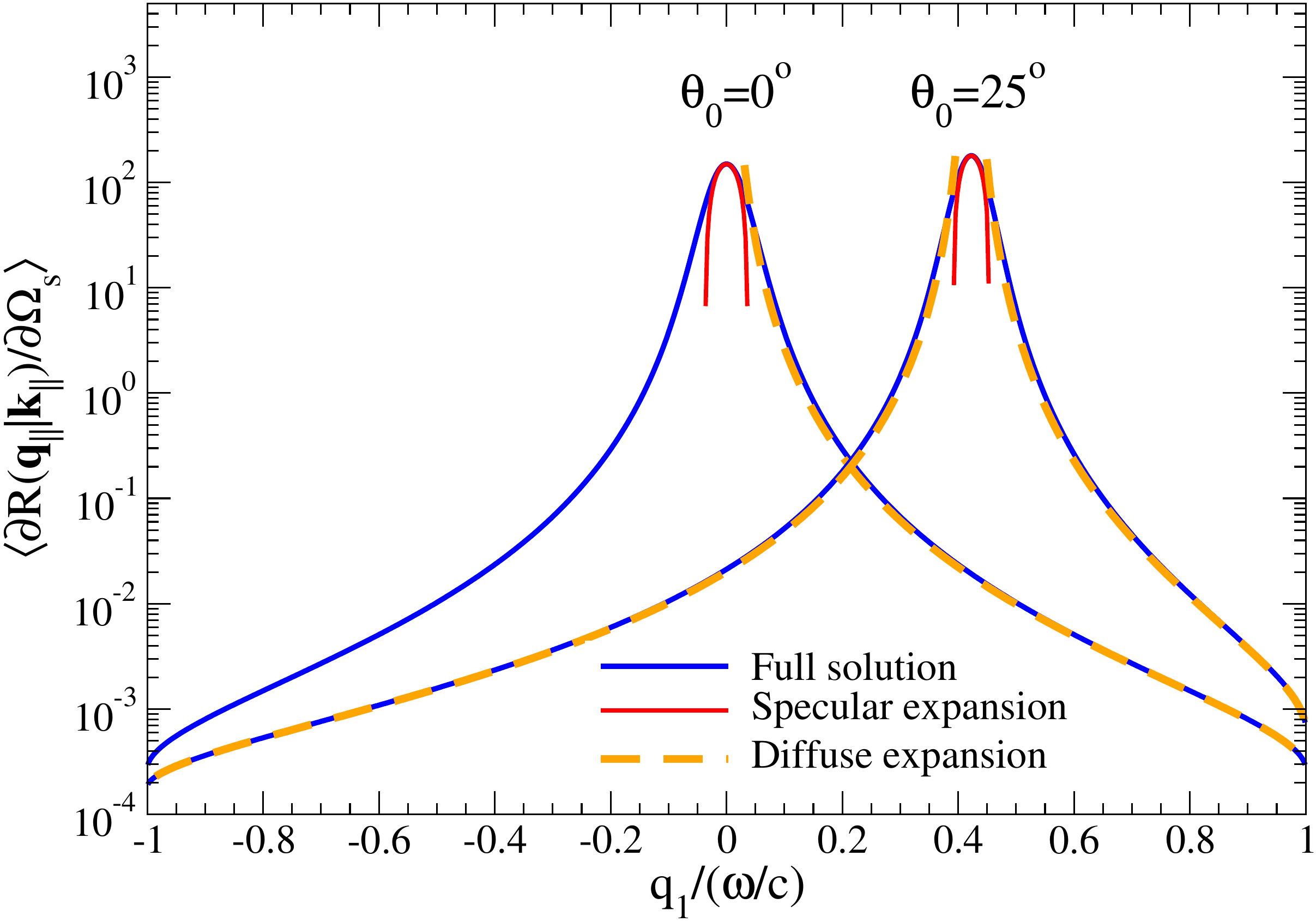}          
  \caption{The mean DRCs calculated on the basis of Eq.~\eqref{eq:MDRC-Kirchoff-Final}[``Full solution'']; Eq.~\eqref{eq:MDRC-specular-expansion} [``Specular expansion'']; and Eq.~\eqref{eq:MDRC-diffuse-expansion} [``Diffuse expansion''] for the polar angles of incidence $\theta_0=\ang{0}$ and \ang{25}. The self-affine surface parameters were $H=0.70$ and $\ell=\num{e-5}\lambda$.}
  \label{fig:MDRC_and_expansions} 
\end{figure}
%

%
%
Figure~\ref{fig:MDRC_and_expansions} presents as solid lines the in-plane dependence of the mean DRC, Eq.~\eqref{eq:MDRC-Kirchoff-Final}, under the assumption that  waves were incident at polar angles $\theta_0=\ang{0}$ and $\theta_0=\ang{25}$ onto a self-affine surface that was characterized by the parameters 
$H=0.70$ and $\ell=\num{e-5}\lambda$. These results display no well-defined specular peaks, that for plane incident waves, should be proportional to $\delta(\pvec{q}-\pvec{k})$ where $\pvec{k}$ is the lateral wave vector of the incident wave. Instead one finds that the self-affine surface gives rise to fully diffuse, wide-angular intensity distributions that are centered around the specular direction. Furthermore, to test the quality of the specular and diffuse expansions, Eqs.~\eqref{eq:MDRC-specular-expansion} and \eqref{eq:MDRC-diffuse-expansion}, respectively, we in Fig.~\ref{fig:MDRC_and_expansions}  also present these expansions. It is found that the quality of both these expansions is rather good, in particular this is the case for the diffuse expansion. The specular expansion is accurate only within a rather narrow region around the specular directions; including higher order terms into the expansion~\eqref{eq:MDRC-specular-expansion} may have extended the region of validity, but here we opted for not doing so.  It should be noted that the quality of the expansions are good for any value of the topothesy (results not shown).

%
%
\smallskip
Before proceeding, one ought to comment on how the mean DRC-data presented in Fig.~\ref{fig:MDRC_and_expansions} (and later figures) were obtained from Eq.~\eqref{eq:MDRC-Kirchoff-Final}. The challenging part of such calculations is the numerical evaluation of the L\'evy distribution defined in Eq.~\eqref{eq:Levy} or equivalent by the Hankel transform form~\eqref{eq:Levy-Hankel}. The Bessel function $\operatorname{J}_0(Q_\parallel v_\parallel)$  that is present in the integral of the latter equation, has an oscillatory character that may lead to loss of numerical significance in the calculation of the integral if not treated properly. To reduce this effect, we performed the calculation in the following way. First we calculated the zeros of the Bessel function $\operatorname{J}_0$.
Then the integration over the original domain was converted into a sum of definite integrals $I_n$ performed between the zeros of the Bessel function.  These definite integrals were evaluated using 
a global adaptive $21$-point Gauss-Kronrod quadrature as implemented in the routine \textsc{qags} from \textsc{QuadPack}~\cite{Book:Doncker-Kapenga1983}. To obtain the final result for the integral present in Eq.~\eqref{eq:Levy-Hankel}, Wynn's $\epsilon$-method~\cite{Wynn1956,Book:Hamming1987} was applied to the sum over $I_n$ for the purpose of improving the convergence rate of the sum (or series). For this purpose the routine \textsc{qelg} from  \textsc{QuadPack} was used. Performing the numerical calculations of the L\'evy distribution in the manner outlined above was found to produce reliable results for all the values of the Hurst exponent and the topothesy that we considered.

\section{Results and Discussions}
\label{Sec:Results} 

This section starts with a presentation and discussion of the properties and features of the expression in Eq.~\eqref{eq:MDRC-Kirchoff-Final} for the mean DRC obtained within the Kirchhoff approximation~[Sec.~\ref{Sec:Result-A}]. Next, the prediction of this expression is compared to what can be obtained for the same quantity from rigorous computer simulations~[Sec.~\ref{Sec:Result-B}]. Finally, in Sec.~\ref{Sec:Result-C} we discuss and give several examples of how the analytic expression for the mean DRC in Eq.~\eqref{eq:MDRC-Kirchoff-Final} can be used together with scattering data for the purpose of reconstructing the self-affine parameters of the surface.

%
%
\begin{figure*}[tb!]
  \centering
  \includegraphics[width=\textwidth]{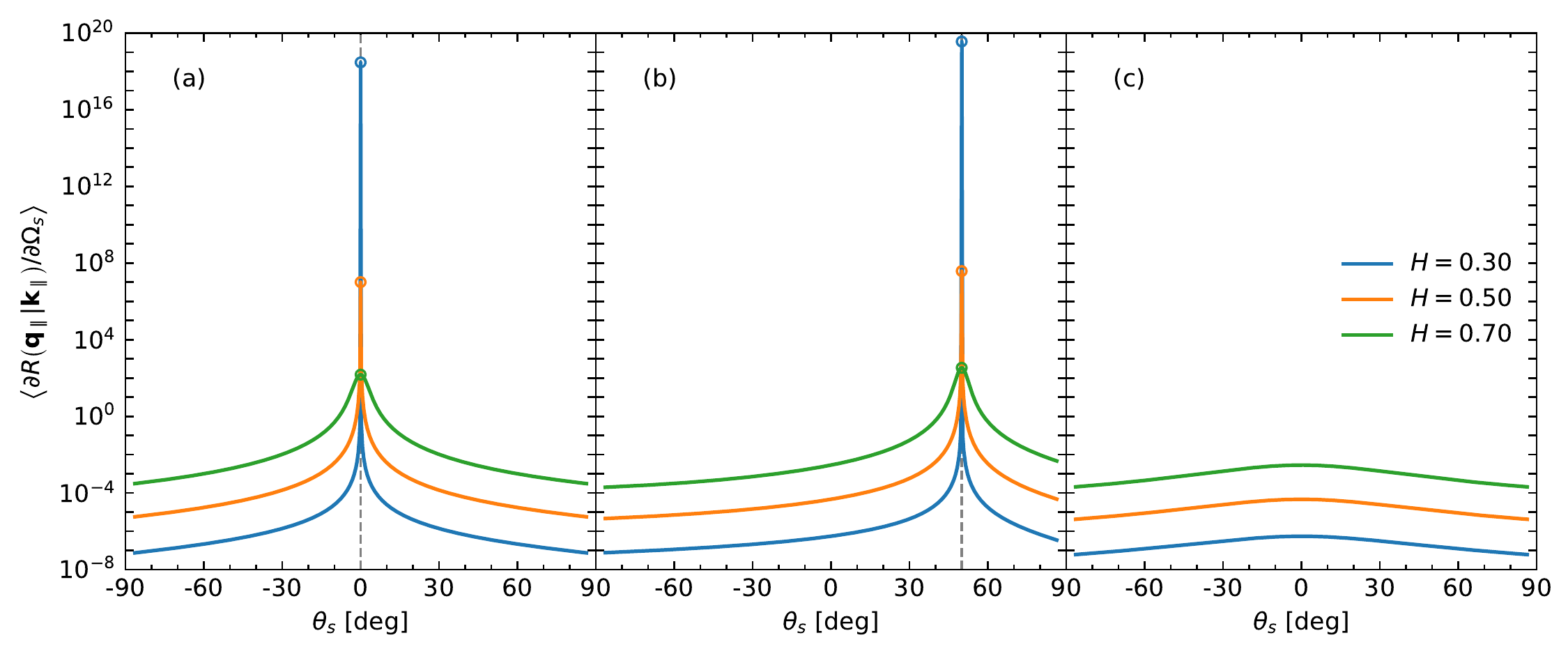}
  \caption{The in-plane or out-of-plane angular dependence of the mean DRC of the scattered wave for self-affine surfaces characterized by Hurst exponents $H=0.30$, \num{0.50} and \num{0.70} and fixed topothesy $\ell=\num{e-5}\lambda$ where $\lambda$ denotes the wavelength of the incident plane wave. The results were obtained on the basis of the analytic expression in Eq.~\eqref{eq:MDRC-Kirchoff-Final}. The subplots correspond to (a) the polar angle of incidence $\theta_0=\ang{0}$ (in-plane and out-of-plane scattering coincide here due to the assumed isotropy of the surface); (b) $\theta_0=\ang{50}$, in-plane scattering; and (c) $\theta_0=\ang{50}$, out-of-plane scattering [$\phi_s=\phi_0\pm\ang{90}$]. In all cases the azimuthal angle of incidence was $\phi_0=\ang{0}$.  The vertical dashed lines in Figs.~\protect\ref{fig:analytical_hurst_comparison}(a)--(b) indicate the specular direction. The open symbols, added for reasons of clarity, represent the value of the mean DRCs at the specular directions. Note the logarithmic scale used on the 2nd axis.} 
  \label{fig:analytical_hurst_comparison}
\end{figure*}
%

\subsection{The mean DRC within the Kirchhoff approximation, Eq.~\eqref{eq:MDRC-Kirchoff-Final}}
\label{Sec:Result-A}
\begin{figure*}[tb!]
  \centering
  \includegraphics[width=\textwidth]{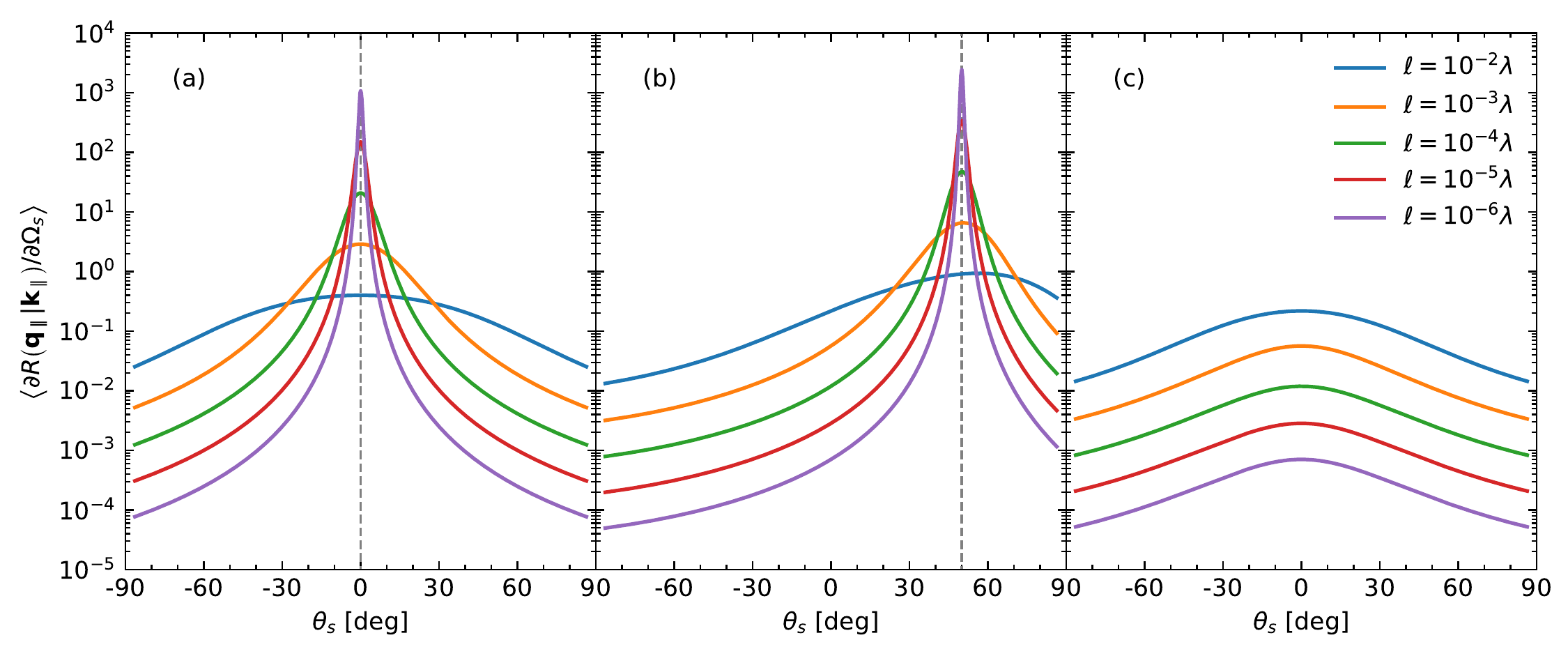}
  \caption{The in-plane or out-of-plane angular dependence of the mean DRC of the scattered wave for self-affine surfaces characterized by  Hurst exponent $H=0.70$ and different values of the topothesy $\ell$. The results were obtained on the basis of the analytic expression in Eq.~\eqref{eq:MDRC-Kirchoff-Final}. The subplots correspond to (a) the polar angle of incidence $\theta_0=\ang{0}$ (in-plane and out-of-plane scattering coincide here due to the assumed isotropy of the surface); (b) $\theta_0=\ang{50}$, in-plane scattering; and (c) $\theta_0=\ang{50}$, out-of-plane scattering [$\phi_s=\phi_0\pm\ang{90}$]. In all cases the azimuthal angle of incidence was $\phi_0=\ang{0}$. The vertical dashed lines in the top two subfigures correspond to the specular direction. Note the logarithmic scale used on the 2nd axis.} 
  \label{fig:analytical_topothesy_comparison}
\end{figure*}

%
%
In Fig.~\ref{fig:analytical_hurst_comparison} we present the in-plane and out-of-plane dependencies of the mean DRCs obtained on the basis of Eq.~\eqref{eq:MDRC-Kirchoff-Final}. These results assumed  three values of the Hurst exponent, $H=0.30$, \num{0.50} and \num{0.70}, and polar angles of incidence $\theta_0=\ang{0}$ and \ang{50}. Without loss of generality, these results were obtained under the assumption that $\phi_0=\ang{0}$ so the $x_1x_3$-plane corresponds to the plane of incidence. The value of the topothesy assumed in producing these results was $\ell=\num{e-5}\lambda$, where $\lambda=2\pi c/\omega$ denotes the wavelength of the incident plane wave. Figure~\ref{fig:analytical_topothesy_comparison} presents the in-plane and out-of-plane dependencies of the mean DRCs for five values of the topothesy  in the range from $\num{e-6}\lambda$ to $\num{e-2}\lambda$ when the value of the Hurst exponent is $H=0.70$.  The results in Figs.~\ref{fig:analytical_hurst_comparison} and \ref{fig:analytical_topothesy_comparison} show that the scattered intensity distributions are all centered around the  specular direction, as expected, which is indicated by the vertical dashed lines in these figures. However, the most striking features of the results presented in Figs.~\ref{fig:analytical_hurst_comparison} and \ref{fig:analytical_topothesy_comparison} are the strong dependencies of the amplitudes and widths of the scattered intensity distributions on the values of the Hurst exponent and the topothesy --- that is, on the parameters characterizing the self-affine surface. For instance, for the situation plotted in Fig.~\ref{fig:analytical_hurst_comparison}(a), one observes that the mean DRC curve corresponding to $H=0.30$ is \num{16} orders of magnitude higher at the specular direction  than the corresponding mean DRC curve for $H=0.70$ for the same direction. A general trend is found in the results reported in Figs.~\ref{fig:analytical_hurst_comparison} and \ref{fig:analytical_topothesy_comparison}; on increasing the value of the Hurst exponent and/or the topothesy, the amplitude of the peak will \textit{decrease} while its width will \textit{increase}. At the same time, the scattered intensity in the tails of the distribution will increase.  For normal incidence, the out-of-plane [$\phi_s=\pm\ang{90}$] scattered intensity distributions are identical to the corresponding in-plane distributions since in this case the distributions are rotational symmetric about the specular direction [Figs.~\ref{fig:analytical_hurst_comparison}(a) and \ref{fig:analytical_topothesy_comparison}(a)]. For non-normal incidence [$\theta_0\neq\ang{0}$], however, this is no longer the case. In Figs.~\ref{fig:analytical_hurst_comparison}(c) and \ref{fig:analytical_topothesy_comparison}(c) we present the out-of-plane dependence of the mean DRCs for a selection of values of the self-affine parameters $H$ and $\ell$. As is to be expected, these results show a symmetry with respect to the plane of incidence and the intensity of the scattered wave is reduced for directions away from $\theta_s=\ang{0}$.

%
%
%

The features observed in Figs.~\ref{fig:analytical_hurst_comparison} and \ref{fig:analytical_topothesy_comparison} concerning the dependence of the mean DRC around the specular direction and how they depend on the self-affine parameters $H$ and $\ell$, can be understood theoretically in terms of the expressions in Eqs.~\eqref{eq:MDRC-specular-direction} and \eqref{eq:peak-width}.
From a more physical perspective, the behavior around the specular direction of the mean DRC can alternatively be understood in terms of the rms-roughness of the surface; for a surface of edges $L$, the global rms-roughness of the surface is $\sigma(L)=\ell(L/\ell)^H$, according to Eq.~\eqref{eq:rms-deviation}. For the situation we are dealing with, $L/\ell \gg 1$, so the rms-roughness of the surface will decrease with decreasing values of $H$ and $\ell$.  In other words, decreasing the values of the self-affine parameters $H$ and $\ell$ will result in a surface that scatters the wave in a more ``mirror-like'' fashion. This is consistent with what is observed from the results of Figs.~\ref{fig:analytical_hurst_comparison} and \ref{fig:analytical_topothesy_comparison}.

\begin{figure}[tb!]
  \centering
  \includegraphics[width=0.9\columnwidth]{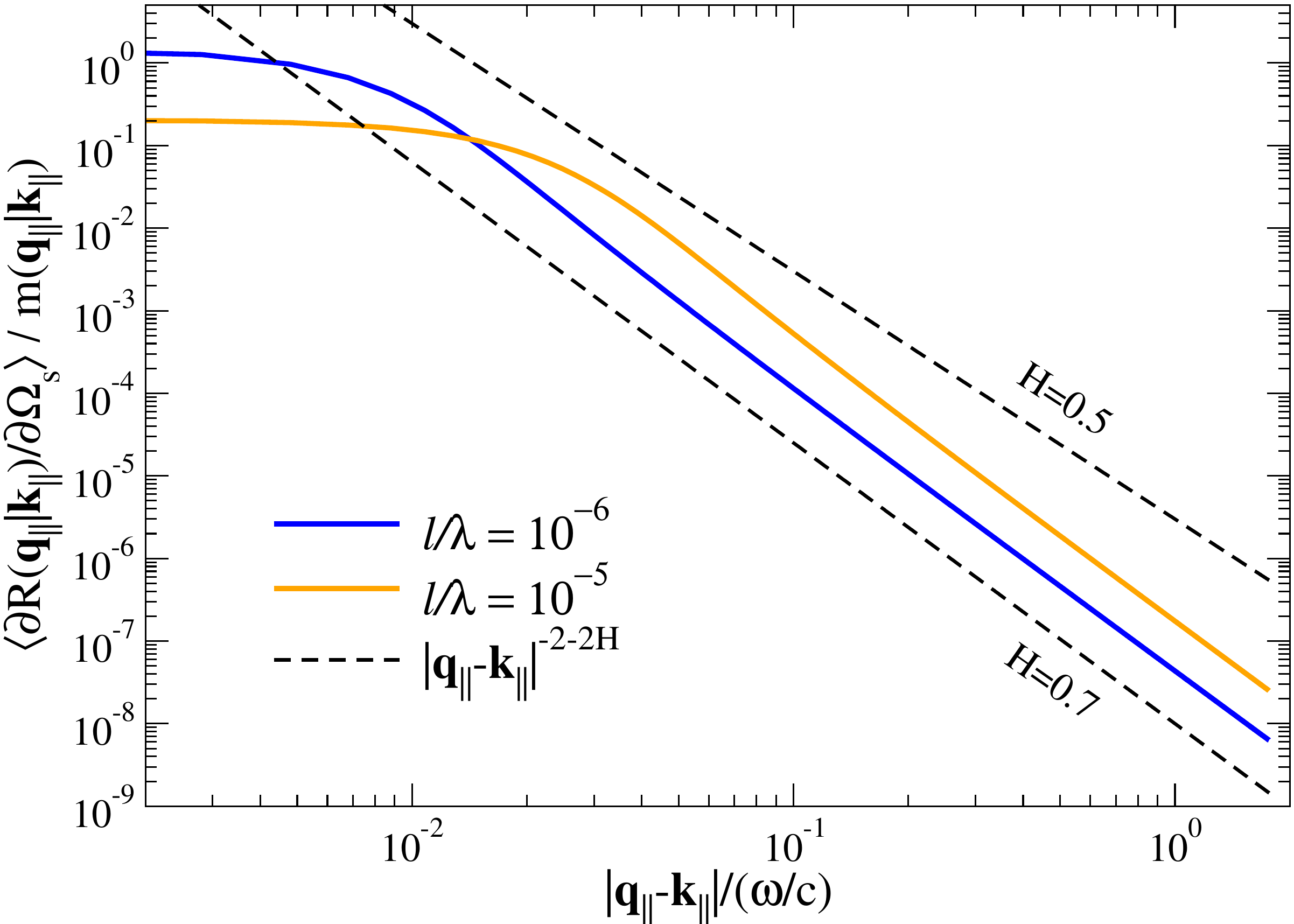}
  \caption{The scaled in-plane mean DRC, $\left< \partial R( \pvec{q} | \pvec{k}) / \partial \Omega_s \right>/ m(\pvec{q}|\pvec{k})$, defined by  Eqs.~\eqref{eq:MDRC-Kirchoff-Final} and \eqref{eq:MDRC-diffuse-expansion-b}, as functions of $|\pvec{q}-\pvec{k}|/(\omega/c)$ using a double logarithmic scale. The self-affine parameters were assumed to $H=0.70$, $\ell=\num{e-6}\lambda$ (blue line) and  $\ell=\num{e-5}\lambda$ (orange line) and the polar angle of incidence was $\theta_0=\ang{50}$.  The dashed lines represent inverse power-law function scaling as $|\pvec{q}-\pvec{k}|^{-2-2H}$ [see Eq.~\eqref{eq:MDRC-diffuse-expansion-a}] for the two values of the Hurst exponent $H=0.70$ and $H=0.50$ as indicated in the figure.} 
  \label{fig:analytical_tail}
\end{figure}

%
%
In the results presented in Figs.~\ref{fig:analytical_hurst_comparison} and \ref{fig:analytical_topothesy_comparison} the inverse power-law tail of the scattered intensity distribution predicted by Eq.~\eqref{eq:MDRC-diffuse-expansion} is not very apparent. To this end, we in Fig.~\ref{fig:analytical_tail} present the in-plane mean DRCs, normalized by the prefactor $m(\pvec{q}|\pvec{k})$ defined in Eq.~\eqref{eq:MDRC-diffuse-expansion-b}, as functions of $|\pvec{q}-\pvec{k}|/(\omega/c)$ for the polar angle of incidence $\theta_0=\ang{50}$ and under the assumption that $H=0.70$ and  $\ell=\num{e-6}\lambda$ or $\ell=\num{e-5}\lambda$. From the results presented in this figure the inverse power-law behavior $|\pvec{q}-\pvec{k}|^{-(2+2H)}$ of the scattered intensity is readily observed in the tail of the distributions. One should in particular note how different values of the Hurst exponent affects the fatness of the tail of the intensity distribution; this is exemplified in Fig.~\ref{fig:analytical_tail} where the dashed lines correspond to the  behavior of the tails of $\left<\partial R /\partial\Omega_s \right>/m(\pvec{q}|\pvec{k})$ 
for Hurst exponents $H=0.70$ and $H=0.50$. As will be demonstrated explicitly below, this dependence can be used to extract the Hurst exponent from scattering data. The results of Figs.~\ref{fig:analytical_topothesy_comparison} and \ref{fig:analytical_tail}, as well as the expression in Eq.~\eqref{eq:MDRC-diffuse-expansion}, also show that the topothesy of the self-affine surface alters the amplitude of the inverse power-law but not its tail exponent; in particular,  Eq.~\eqref{eq:MDRC-diffuse-expansion} predicts that the amplitude of the tail should scale with the topothesy as $\ell^{2-2H}$.

%
%
%

\begin{figure*}[tb!]
  \centering
  \includegraphics[width=0.8\textwidth]{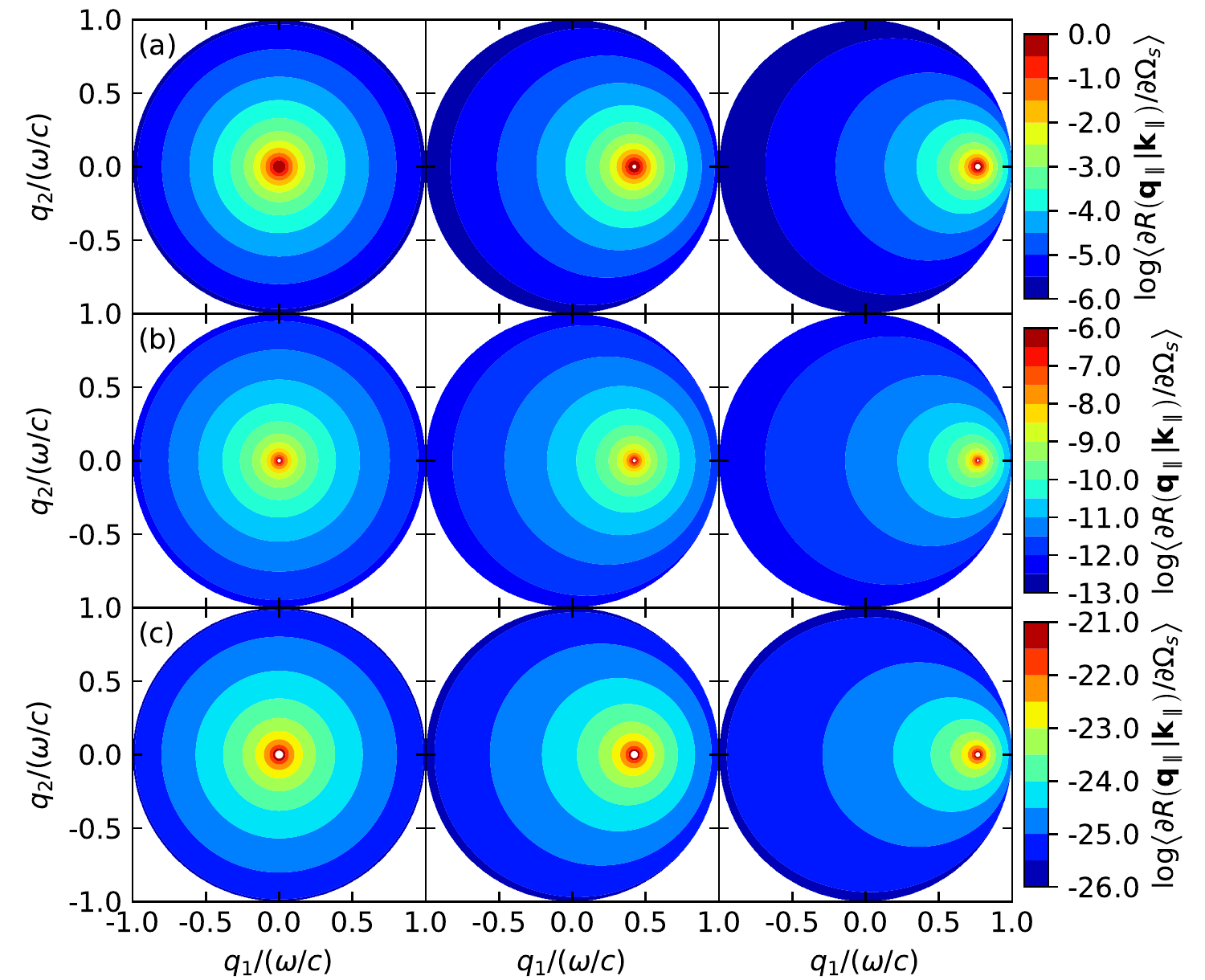}
  \caption{
    The full angular distribution of the mean DRC of the scattered wave based on Eq.~\eqref{eq:MDRC-Kirchoff-Final} for self-affine surfaces for a series of Hurst exponents and incident angles. The topothesy is fixed at $\ell/\lambda=10^{-5}$ where $\lambda$ denotes the wavelength of the incident plane wave. Row (a) corresponds to $H=0.70$, row (b) to $H=0.50$, and row (c) to $H=0.30$. The polar angle of incidence is constant in each column, and is $\theta_0=\ang{0}$, \ang{25} and \ang{50} for each column from left to right. Notice the significant difference in values of the color scale between the rows.} 
  \label{fig:analytical_full_angular} 
\end{figure*}

%
%
Up till now we have presented either in-plane or out-of-plane cuts of the scattered intensity distributions. It is instructive to also have available the full angular intensity distribution of the scattered intensity. Therefore, in Fig.~\ref{fig:analytical_full_angular} we present contour plots of the angular dependence of the logarithm of the mean DRCs for Hurst exponents $H\in\{0.70, 0.50, 0.30\}$ [rows of sub-figures in Fig.~\ref{fig:analytical_full_angular}], topothesy $\ell=\num{e-5}\lambda$ and a set of angles of incidence $\theta_0\in\{\ang{0}, \ang{25}, \ang{50}\}$ [columns in Fig.~\ref{fig:analytical_full_angular}]. The results of this figure show, as is to be  expected, that the scattered intensity distributions are rotational symmetric when the wave is incident normally onto the self-affine surface. For non-normal incidence, $\theta_0\neq \ang{0}$, only sufficiently close to the specular direction do we observe an approximate rotational symmetry around this direction. However, as we move away from the specular direction the rotation symmetry about the specular direction is lost, while a mirror symmetry with respect to the plane-of-incidence, the $q_2q_3$-plane,  remains.

%
%
%

\begin{figure*}[tb!]
  \centering
  \includegraphics[width=\textwidth]{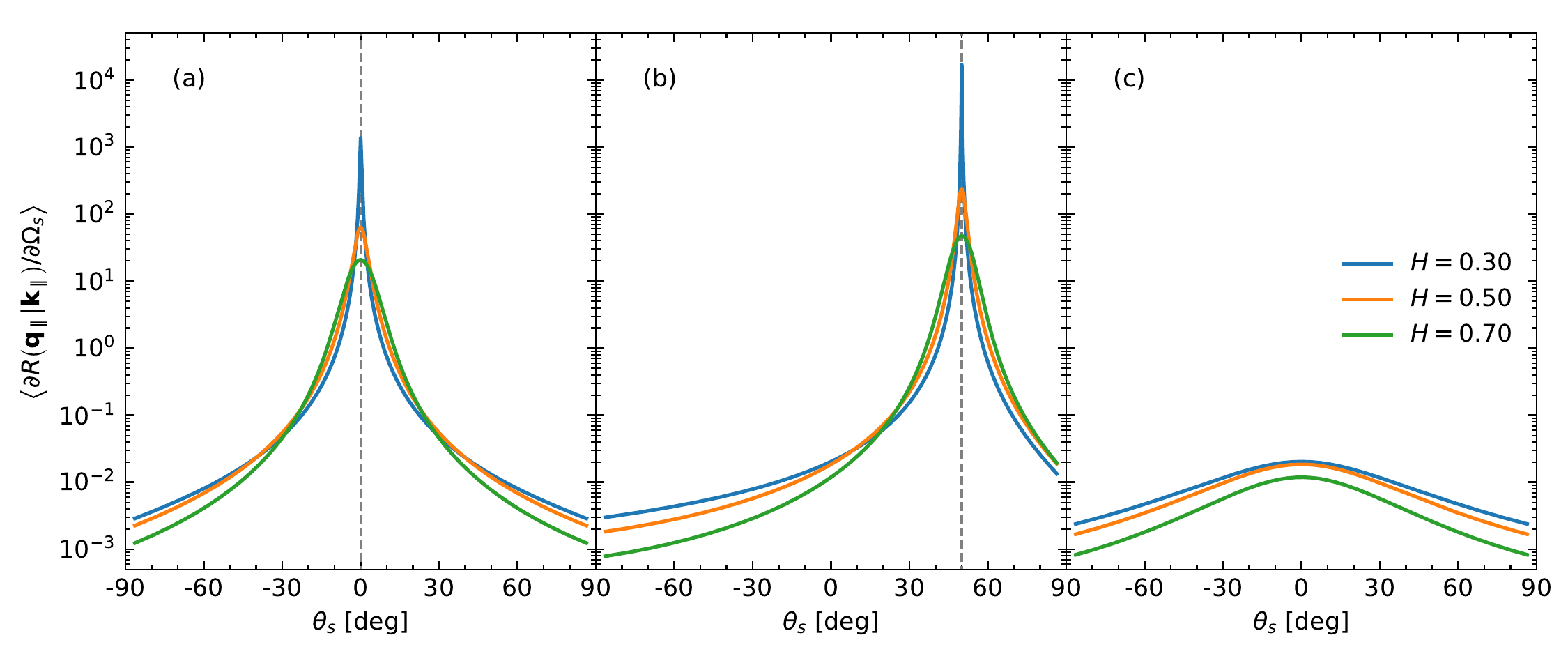}
  \caption{The in-plane or out-of-plane dependence of the mean DRC of the scattered wave for self-affine surfaces characterized by the same slope $s(\lambda)=0.0631$. The results were obtained on the basis of the analytic expression in Eq.~\eqref{eq:MDRC-Kirchoff-Final}. The subplots correspond to the polar angle of incidence (a) $\theta_0=\ang{0}$ (in-plane and out-of-plane scattering coincide here due to the assumed isotropy of the surface); (b) $\theta_0=\ang{50}$, in-plane scattering; and (c) $\theta_0=\ang{50}$, out-of-plane scattering. The vertical dashed lines in panels~(a) and (b) correspond to the specular direction. Note the logarithmic scale used on the 2nd axis.}
  \label{fig:analytical_constant_slope} 
\end{figure*}

%
%
\smallskip
In the studies of wave scattering from one-dimensional self-affine surfaces~\cite{Simonsen1999-3}, it was argued that the rms-slope over a wavelength $s(\lambda)$, Eq.~\eqref{eq:rms-slope}, potentially  is a more relevant parameter to characterize the angular dependent intensity scattered from self-affine surfaces than the topothesy; this is in particular the case when comparing the scattering from surfaces of different Hurst exponents. As the reader can confirm, the expressions in, for instance, Eqs.~\eqref{eq:MDRC-Kirchoff-Final} and \eqref{eq:peak-width}  can readily be expressed in terms of $s(\lambda)$ --- the average slope over a lateral distance that equals the wavelength of the incident wave --- as was the case for the corresponding result for the one-dimensional self-affine surfaces. In Fig.~\ref{fig:analytical_constant_slope} we compare the angular dependent in-plane and out-of-plane variation of the mean DRCs for self-affine surfaces characterized by the rms-slope $s(\lambda)=0.0631$ and different Hurst exponents. These results show that when the averages slope of the surface is constant, significantly less variation with Hurst exponent is observed for the intensity scattered into the specular direction, as compared to, for instance,  the results depicted in  Fig.~\ref{fig:analytical_hurst_comparison} for which the topothesy was kept constant.
This indicates, as was pointed out for one-dimensional self-affine surfaces in Ref.~\onlinecite{Simonsen1999-3}, that the rms-slope over a wavelength is a relevant quantity for characterizing the scattered intensity from a self-affine surface.



%
%
%

%
%
\subsection{Comparison to rigorous computer simulation results}
\label{Sec:Result-B}

\begin{figure}[tb!]
  \centering
  \includegraphics[width=\columnwidth]{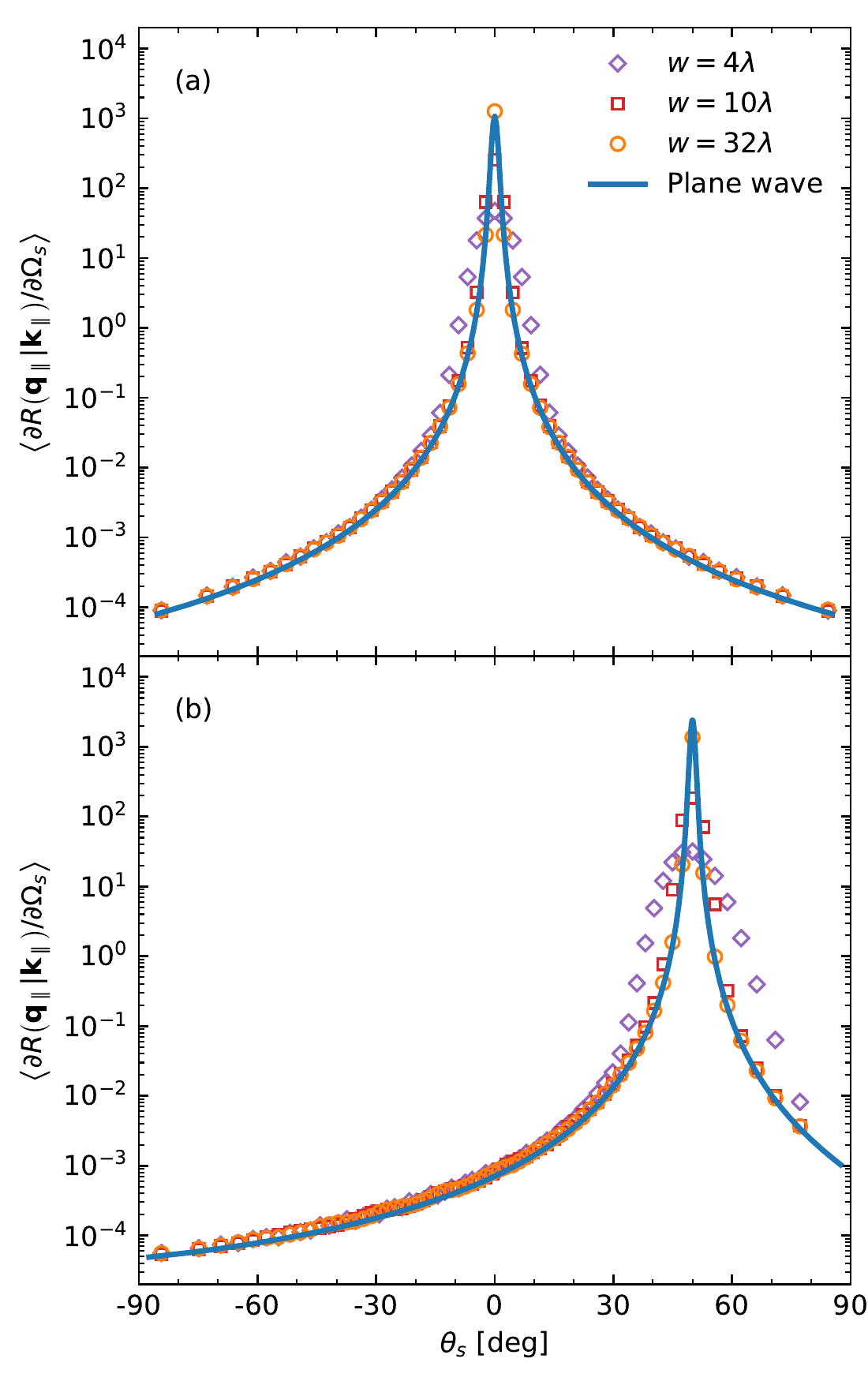}
  \caption{\label{fig:kirchhoff_analytical_comparison}
    Comparison of the in-plane angular dependence of the mean DRC obtained on the basis of the Kirchhoff approximation using an incident plane wave, Eq.~\eqref{eq:MDRC-Kirchoff-Final}, or a Gaussian beam of width $w/\lambda\in\{4,10,32\}$, for a elf-affine surface of parameters  $H=0.70$ and $\ell=\num{e-6}\lambda$. The polar angles of incidence were (a)~$\theta_0=\ang{0}$ and (b)~$\theta_0=\ang{50}$. The value of the edges of the surfaces assumed in obtaining these results was $L=3w$.}
\end{figure}

The analytic expression for the mean DRC in Eq.~\eqref{eq:MDRC-Kirchoff-Final} was derived within the Kirchhoff approximation, which is a single scattering approximation, and under the assumption that the incident wave is a plane wave. We will now compare the prediction from this expression to the results that can be obtained from rigorous computer simulations which take all multiple scattering effects into account. Such simulations were performed on the basis of Green's second integral identity by the method detailed in Refs.~\cite{Tran1992,Macaskill1993,Tran1993,HardWallPaper}. 
Due to the restrictions on computer resources, like computer memory and cpu-time, we were unfortunately not able to perform simulations for plane wave illumination which would require the surface area covered by the rough surface to be large to suppress diffraction effects from the edges of the surface. Instead an incident finite-sized Gaussian beam was used when performing the rigorous computations~\cite{Tran1993,HardWallPaper} and its use reduces, and potentially eliminates, edge effects. However, the use of an incident finite beam has the undesired side effect that the scattered intensity at, and around, the specular direction will differ from what is obtained when the surface is illuminated by a plane incident wave. An incident Gaussian beam can be modeled as a superposition of plane waves of amplitudes that gradually go to zero for directions away from the intended propagation direction of the beam; see Refs.~\onlinecite{Tran1993} and \onlinecite{HardWallPaper} for details. A sample that produces a scattered intensity distribution displaying a well-defined peak in the specular direction when illuminated by a plane wave is therefore expected to produce a less intense and broader peak around the same direction when illuminated by a Gaussian beam of the same polar angle of incidence. That such behavior indeed is correct is illustrated in Fig.~\ref{fig:kirchhoff_analytical_comparison} where the in-plane dependencies of the mean DRCs are presented for the polar angles of incidence $\theta_0=\ang{0}$ and $\theta_0=\ang{50}$ using the analytic expression~\eqref{eq:MDRC-Kirchoff-Final} 
that assumes a plane incident wave [solid lines], and simulation results performed on the basis of the Kirchhoff approximation using an incident Gaussian beam of width $w/\lambda\in\{4,10,32\}$ [open symbols]. From the different results presented in Fig.~\ref{fig:kirchhoff_analytical_comparison}, it is observed that a plane incident wave produce the most intense and the most narrow peak in the specular direction. On the other hand, the most narrow incident Gaussian beam [$w=4\lambda$] cause the broadest and less intense peak in the specular direction. As the width of the incident Gaussian beam is increased, the scattered intensity distribution that it gives rise to, starts to approach the distribution produced by an incident plane wave. In particular, it is noted from the results in  Fig.~\ref{fig:kirchhoff_analytical_comparison}  that the computer simulation result produced assuming an incident Gaussian beam of width $w = 32\lambda$ and surface edges $L = 96\lambda$, is rather close to the plane wave prediction; at least this was the case for the self-affine parameters assumed in producing this result. For the relevant practical applications that we are concerned about, the width of the incident beam will be much larger than its wavelength, $w\gg \lambda$. In such cases the assumption of a plane incident wave should not represent any serious restriction.
%
It should be mentioned that we could have pursued a derivation of an analytic expression similar to Eq.~\eqref{eq:MDRC-Kirchoff-Final} but which assumes a Gaussian beam as the source of illumination. Here we have not done so for several reasons. First, for most practically relevant cases a plane incident wave is a fair assumption. Second, the mathematical expression obtained for the mean DRC  using a Gaussian beam would have been significantly more complicated without any significant benefit.

%
%
%

\begin{figure*}[tb!]
  \centering
 \includegraphics[width=\textwidth]{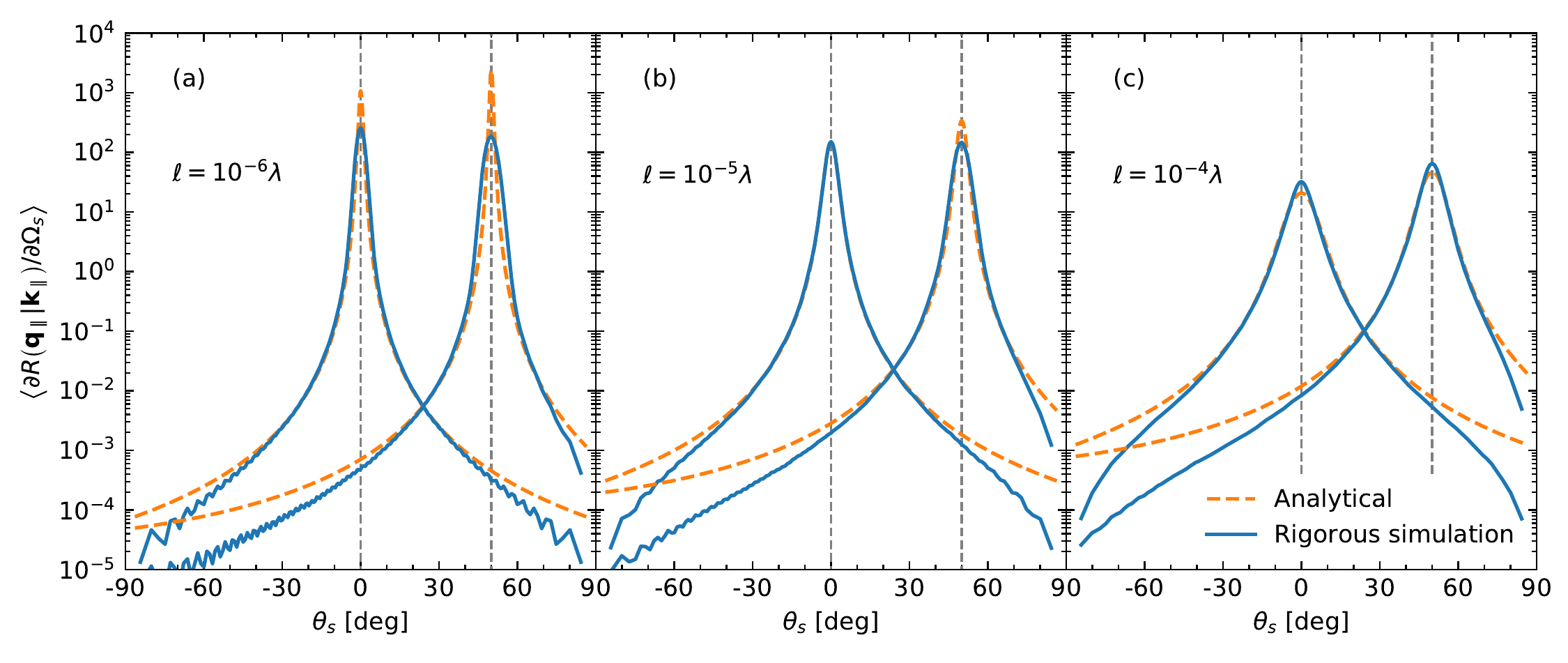}
  \caption{The in-plane dependence of the mean DRC of the scattered wave for self-affine surfaces characterized by Hurst exponent $H=0.70$ and different values of the topothesy $\ell$ using the Dirichlet boundary condition. The wavelength of the incident wave was $\lambda$. The results shown with solid lines were obtained on the basis of rigorous simulations that assumed an incident Gaussian beam of width $w=10\lambda$, and the dashed lines show results based on the analytic expression in Eq.~\eqref{eq:MDRC-Kirchoff-Final} derived under the assumption of an incident plane wave. The subplots correspond to the topothesy: (a) $\ell=10^{-6}\lambda$, (b) $\ell=10^{-5}\lambda$, (c) $\ell=10^{-4}\lambda$.
  Each plot contains the in-plane cuts for polar incidence angle $\theta_0=\ang{0}$ and $\theta_0=\ang{50}$. The vertical dashed lines in panels~(a) and (b) correspond to the specular direction. The vertical dashed lines in panels~(a) and (b) correspond to the specular direction. Note the logarithmic scale used on the 2nd axis.}
  \label{fig:Dirichlet_rigorous_topothesy_comparison} 
\end{figure*}

We now turn to a comparison of the results obtained from the single scattering results in Eq.~\eqref{eq:MDRC-Kirchoff-Final} and the results that can be obtained from rigorous computer simulations for the scattered intensity. The first set of rigorous computer simulation results that we will present are for the wave scattering from self-affine Dirichlet surfaces, and the obtained results are present in Fig.~\ref{fig:Dirichlet_rigorous_topothesy_comparison} as solid lines. This figure shows the in-plane angular dependence of the mean DRCs as functions of the scattering angle $\theta_s$ for self-affine surfaces characterized by $H=0.70$ and $\ell/\lambda\in\{\num{e-4},\num{e-5},\num{e-6}\}$. In producing these and all remaining simulation results in this work the width of the incident Gaussian beam was assumed to be $w=10\lambda$, if nothing is said to indicate otherwise; the value of the edges of the surfaces was $L=3w$; and  the spatial discretization length was assumed to be $\Delta x_\parallel = \lambda/7$.
The reported results were obtained as an  averaging over $N_\zeta=\num{5000}$ surface realizations. These simulation results satisfied energy conservation within an error of  no greater than $0.1\%$. The prediction based on the analytic expression~\eqref{eq:MDRC-Kirchoff-Final}, assuming the same self-affine parameters used in performing the rigorous simulations, are displayed in Fig.~\ref{fig:Dirichlet_rigorous_topothesy_comparison} as dashed lines. The results in  Fig.~\ref{fig:Dirichlet_rigorous_topothesy_comparison} reveal a rather good agreement between the analytic results~\eqref{eq:MDRC-Kirchoff-Final} and the rigorous computer simulation results. The largest discrepancy between these two sets of results are found for the largest scattering angles for which the analytic expression, for a Dirichlet surface, overestimates the value of the scattered intensity. Furthermore, increasing the value of the topothesy also increases the discrepancy, but the increase is not dramatic. It should also be noted from Fig.~\ref{fig:Dirichlet_rigorous_topothesy_comparison} that increasing the topothesy of the surface seems to produce a better agreement between the two sets of results around the specular direction. This we believe is a result of the increased  diffusive nature of the scattered intensity distributions obtained for self-affine surfaces of increasing value of the  topothesy. The fair agreement between the two sets of results depicted in Fig.~\ref{fig:Dirichlet_rigorous_topothesy_comparison}
indicates that multiple scattering effects  do not play any significant role for the self-affine parameters assumed in obtaining these results.
%

%
%
%

\begin{figure*}[tb!]
  \centering
\includegraphics[width=\textwidth]{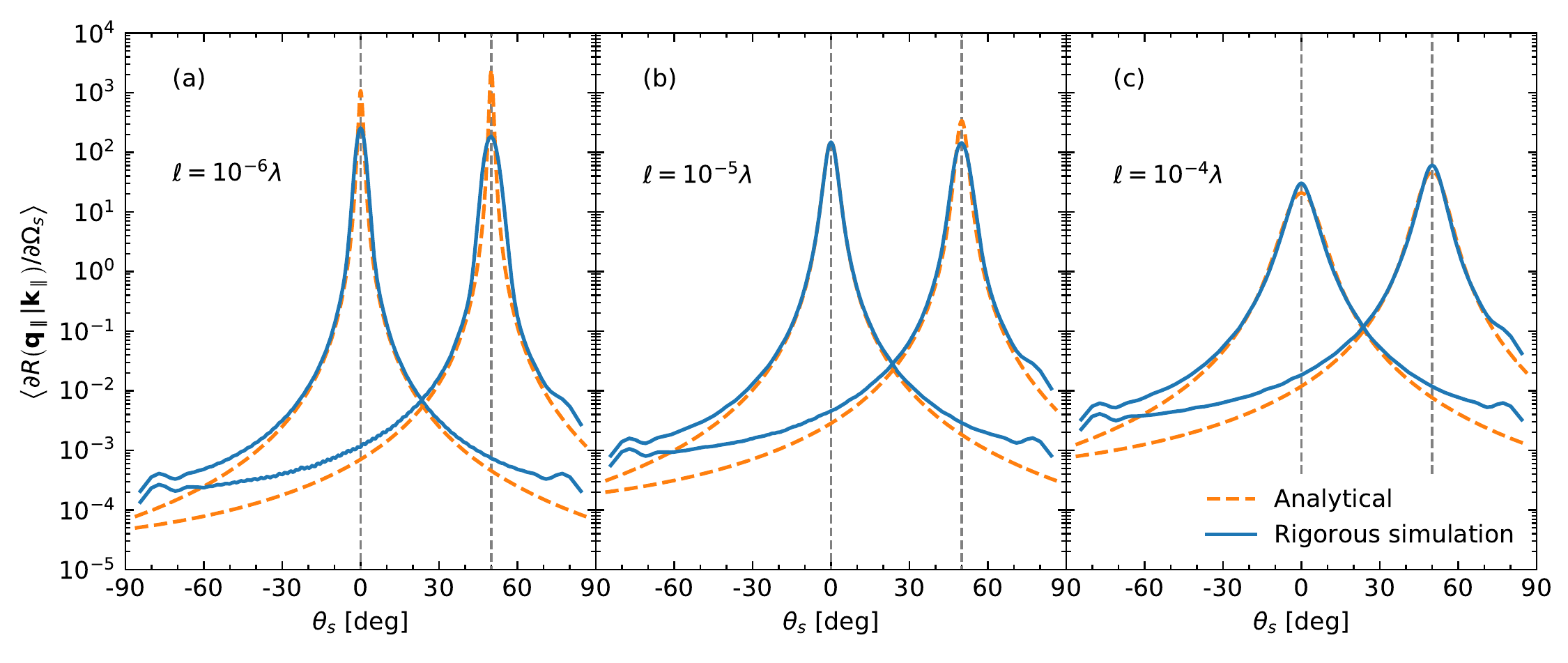}
\caption{Same as Fig.~\protect\ref{fig:Dirichlet_rigorous_topothesy_comparison} but for a self-affine Neumann surface [$H=0.70$].}
  \label{fig:Neumann_rigorous_topothesy_comparison} 
\end{figure*}

Figure~\ref{fig:Neumann_rigorous_topothesy_comparison}  presents the in-plane and out-of-plane intensity distribution scattered from self-affine Neumann surfaces. With the exception of the difference in boundary condition that applies on the rough surface, this is exactly the same scattering system for which the results were presented for Dirichlet surfaces in Fig.~\ref{fig:Dirichlet_rigorous_topothesy_comparison}. We recall that within the Kirchhoff approximation the scattering amplitude for the Neumann and the corresponding Dirichlet system only differ by a sign so that the mean DRC for the two problems are the same, see Eq.~\eqref{eq:MDRC-Kirchoff-Final}. However, when the scattering amplitudes are obtained by rigorous means this is no longer the case. By comparing the simulation results presented in Figs.~\ref{fig:Dirichlet_rigorous_topothesy_comparison} and~\ref{fig:Neumann_rigorous_topothesy_comparison} it is observed that qualitatively the results obtained for the two systems are still rather similar. However, a closer inspection of these results reveals that the scattered intensity in the tails of the intensity distribution is different. While the analytic expression for the mean DRC~\eqref{eq:MDRC-Kirchoff-Final} overestimated the scattered intensity in the tails of the distribution for Dirichlet surfaces the same expression underestimates the intensity for the corresponding Neumann surfaces.

%
%
%

\begin{figure*}[tb!]
  \centering
  \includegraphics[width=\textwidth]{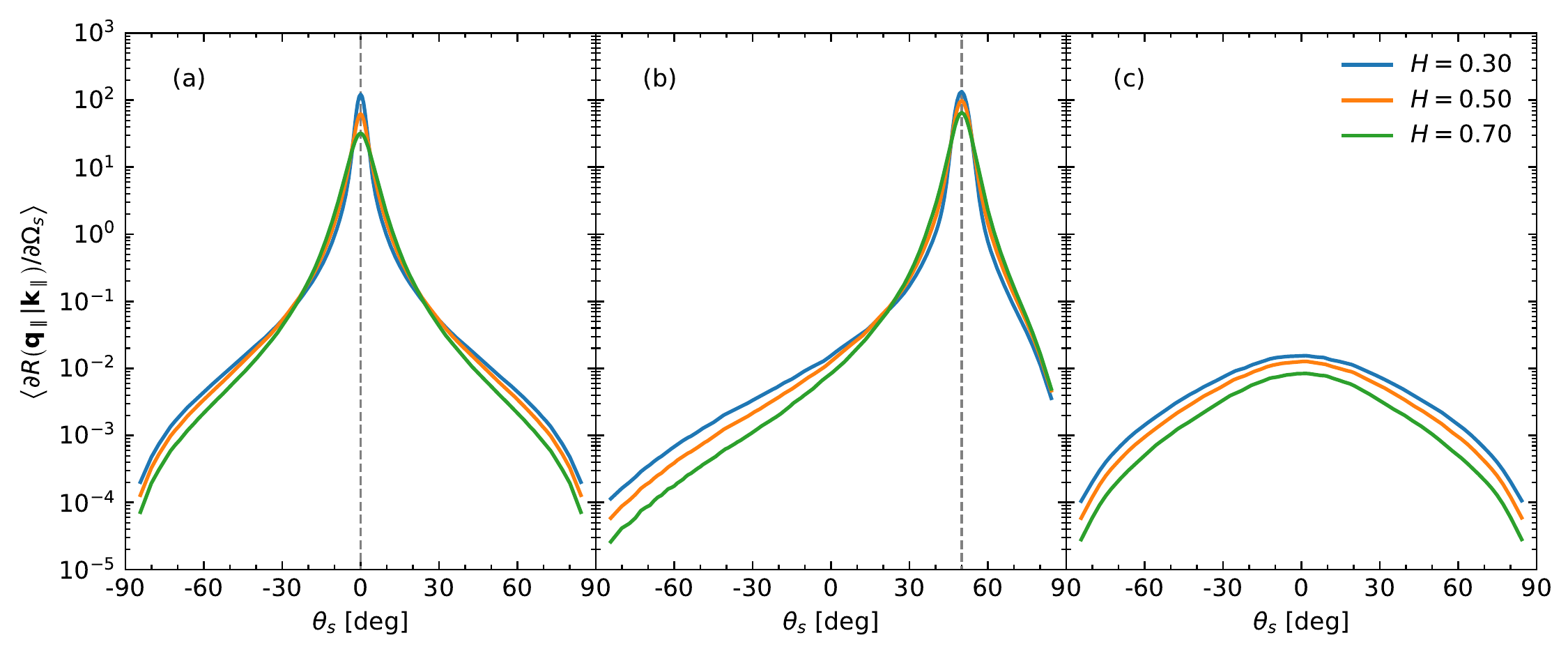}
  \caption{The in-plane or out-of-plane dependence of the mean DRC of the scattered wave for self-affine surfaces characterized by the slope $s(\lambda)=0.0631$. The results were obtained on the basis of rigorous simulations using the Dirichlet boundary condition. The subplots correspond to the polar angle of incidence (a) $\theta_0=\ang{0}$ (in-plane and out-of-plane scattering coincide here due to the assumed isotropy of the surface); (b) $\theta_0=\ang{50}$, in-plane scattering; and (c) $\theta_0=\ang{50}$, out-of-plane scattering. The vertical dashed lines in panels~(a) and (b) correspond to the specular direction. Note the logarithmic scale used on the 2nd axis.} 
  \label{fig:rigorous_constant_slope} 
\end{figure*}

We have also performed rigorous simulation for self-affine surfaces of constant slope over a wavelength, $s(\lambda)$. The results are presented for the Dirichlet boundary condition in Fig.~\ref{fig:rigorous_constant_slope} for slope $s(\lambda)=0.0631$ and the same values for the self-affine parameter assumed in producing Fig~\ref{fig:analytical_constant_slope}. Reasonable quantitative agreement is found between the corresponding results from Figs.~\ref{fig:analytical_constant_slope} and~\ref{fig:rigorous_constant_slope}.

%
%
%

\subsection{Extraction of self-affine parameters from measured scattering data}
\label{Sec:Result-C}
%
%

%
%
\begin{figure*}[tb!]
  \centering
  \includegraphics[height=0.32\textwidth]{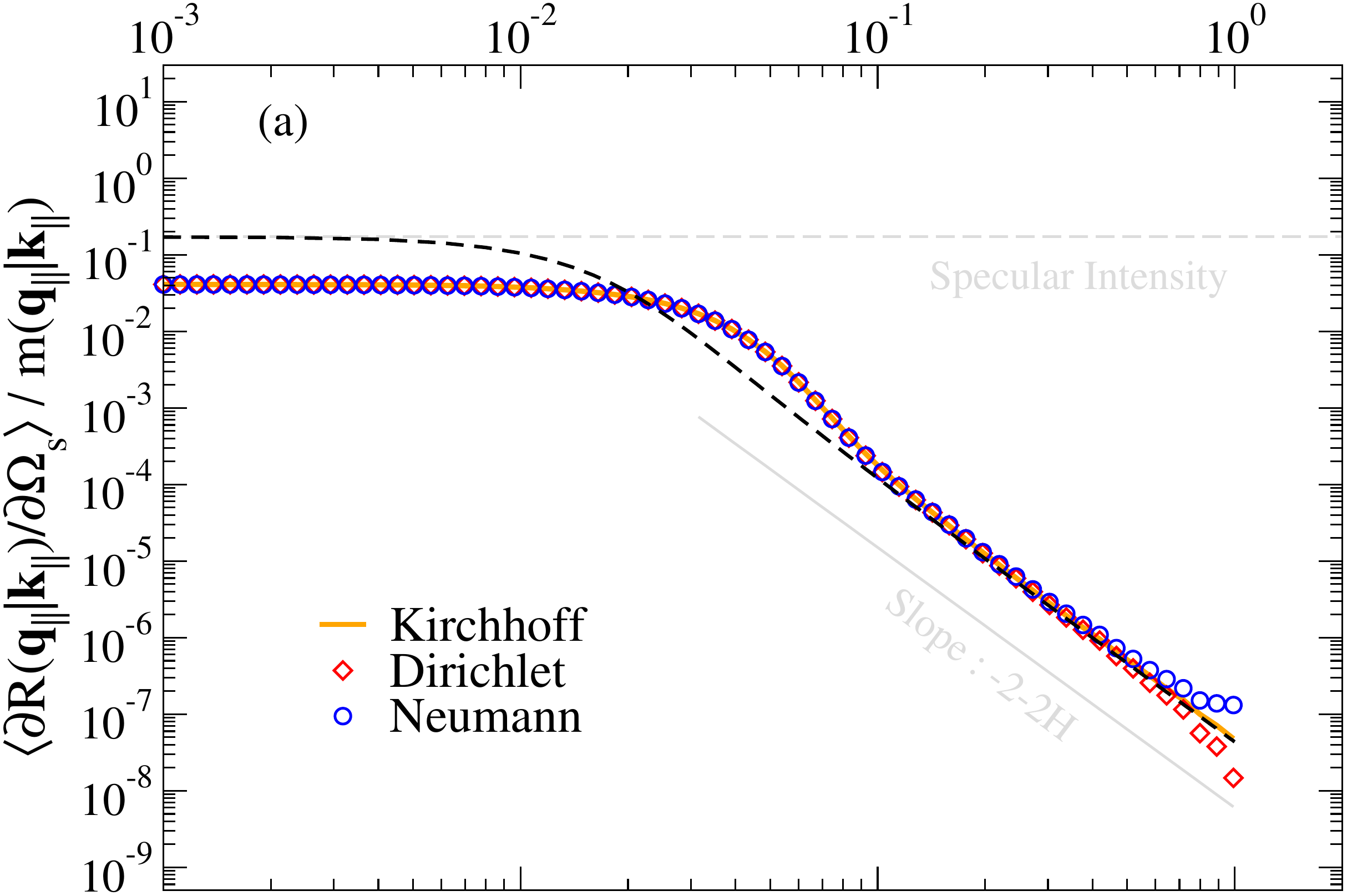}
  \!\! 
  \includegraphics[height=0.32\textwidth]{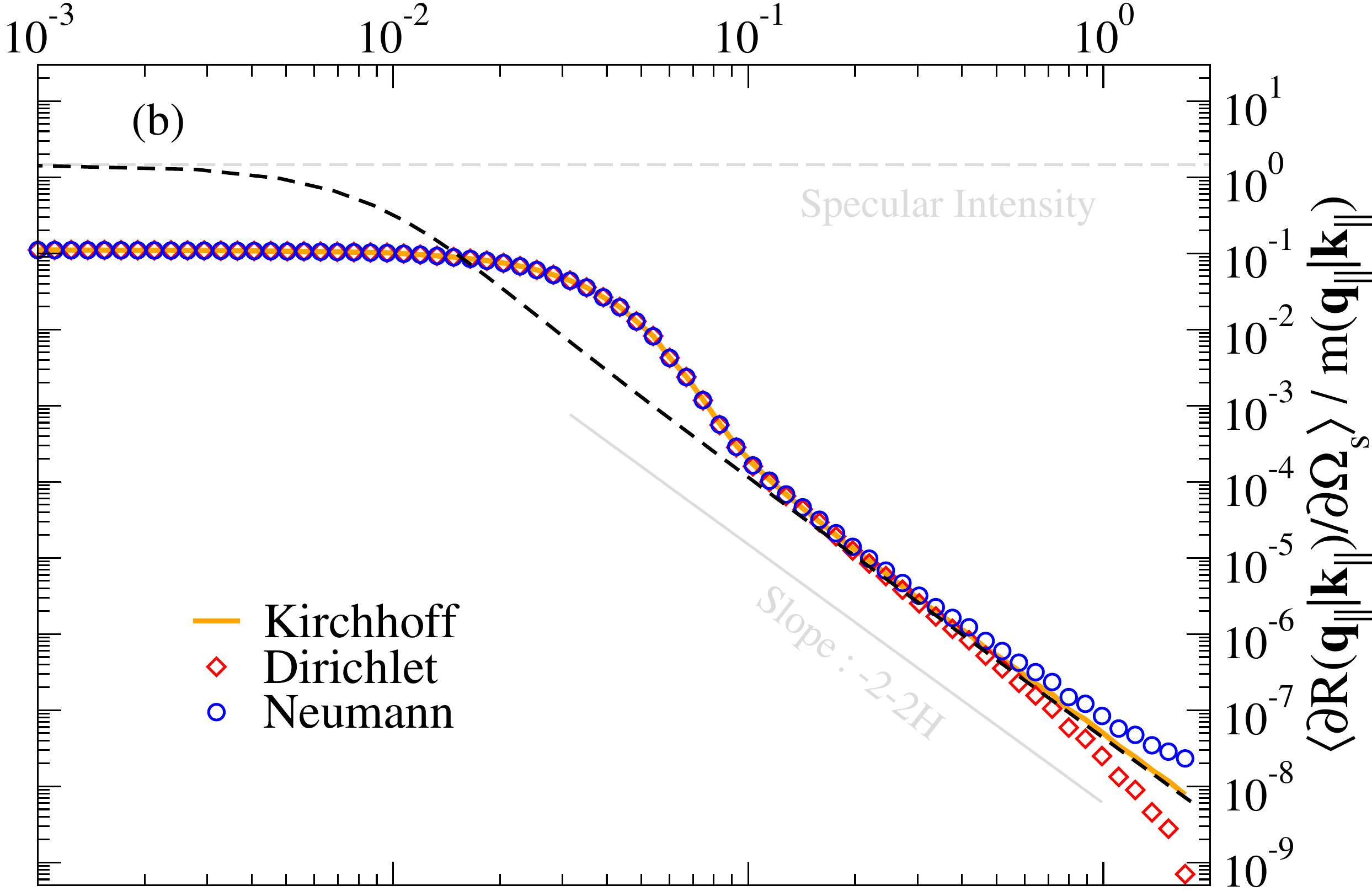}  
  \\*[0.7ex]
  \includegraphics[height=0.341\textwidth]{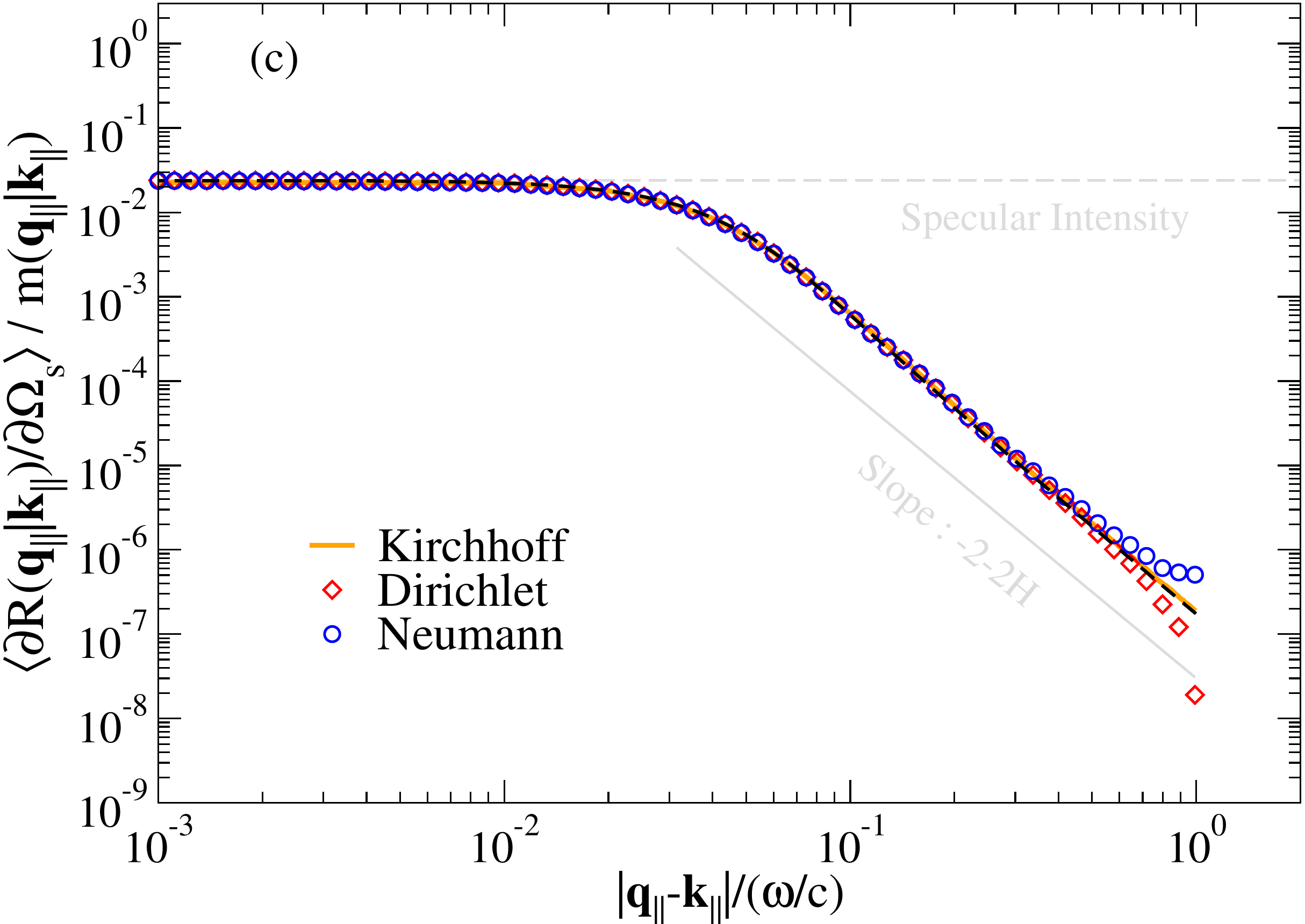}
  \!\! 
  \includegraphics[height=0.341\textwidth]{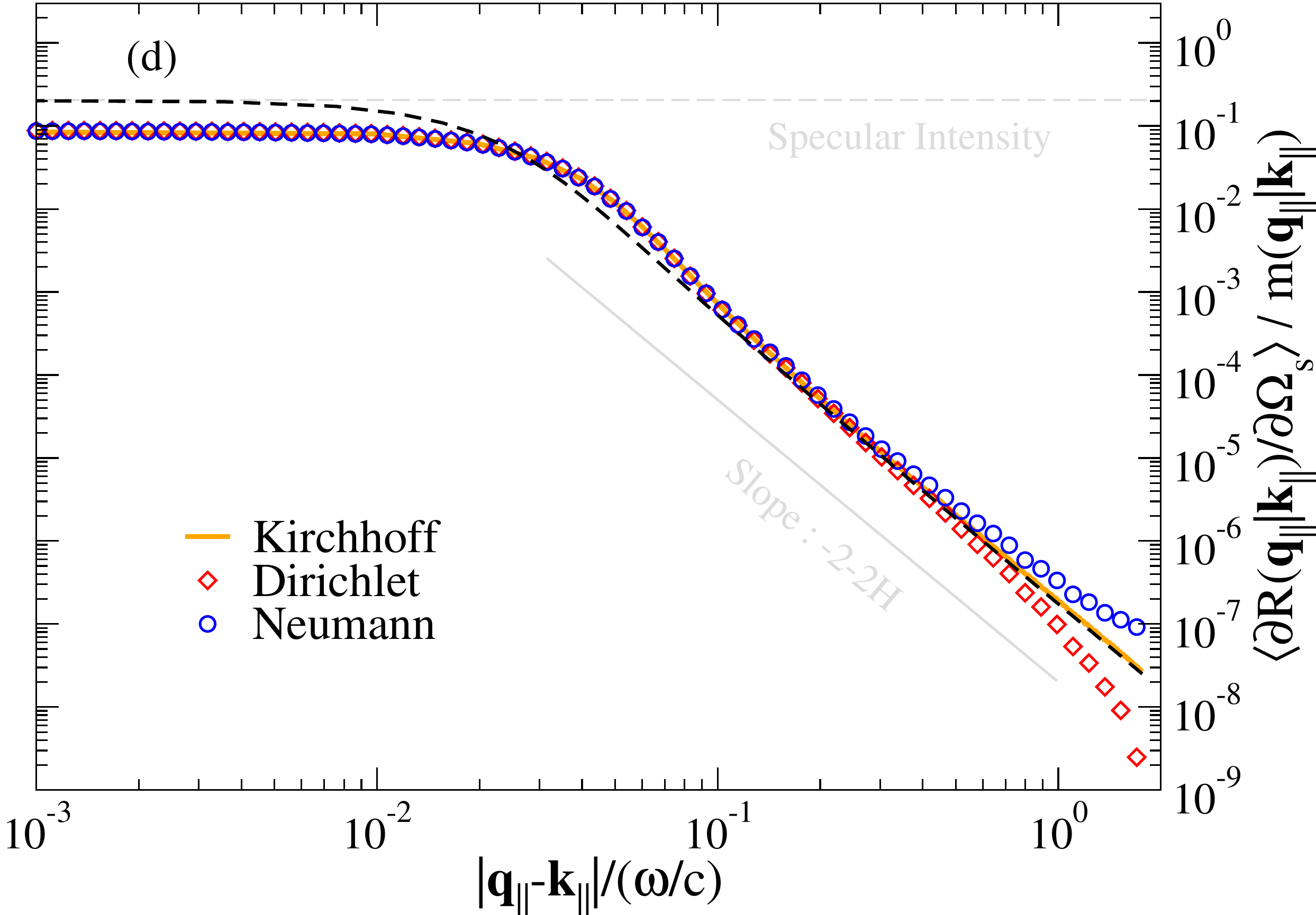}
  \caption{The scaled mean DRCs, $\left< \partial R( \pvec{q} | \pvec{k}) / \partial \Omega_s \right>/ m(\pvec{q}|\pvec{k})$,
    as functions of $|\pvec{q}-\pvec{k}|/(\omega/c)$ obtained within the Kirchhoff approximation~(solid and dashed lines) or by rigorous computer simulations for Dirichlet and Neumann surfaces~(open symbols). The parameters assumed for the self-affine surfaces were $H=0.70$, $\ell=\num{e-6}\lambda$~[Figs.~\protect\ref{fig:Hurst_extraction}(a)--(b)] and
    $\ell=\num{e-5}\lambda$~[Figs.~\protect\ref{fig:Hurst_extraction}(c)--(d)],
the wavelength of the incident wave was $\lambda$ and the polar angles of incidence were $\theta_0=\ang{0}$~[Figs.~\protect\ref{fig:Hurst_extraction}(a,\,c)] and (b)~$\theta_0=\ang{50}$~[Figs.~\protect\ref{fig:Hurst_extraction}(b,\,d)]. The dashed black lines were obtained on the basis of Eq.~\eqref{eq:MDRC-Kirchoff-Final} and therefore assume plane wave illumination. The result corresponding to the solid lines  and rigorous results~(open symbols) were obtained under the assumption of  incident Gaussian beams of width $w=10\lambda$ and the edges of the surfaces were $L=3w=30\lambda$. In performing the rigorous simulations, the sampling interval assumed for the surfaces was $\Delta x_\parallel =\lambda/7$ and the reported results were obtained as averages over $N_\zeta=\num{1000}$ surface realizations. As a guide to the eye we have included two sets of gray lines; the solid gray lines of slopes $-2-2H$ represent the tail behavior of the scattered intensity distribution in Eq.~\eqref{eq:MDRC-diffuse-expansion} and the dashed horizontal gray lines correspond to the specular intensity for a plane incident wave, Eq.~\eqref{eq:MDRC-specular-direction}.}
  \label{fig:Hurst_extraction} 
\end{figure*} 
%

After having established that the expression in Eq.~\eqref{eq:MDRC-Kirchoff-Final} well represents the intensity scattered from a self-affine surface, we now turn to how it in combination with the results of a scattering experiment performed on a self-affine surface can be used to determine the Hurst exponent and potentially the topothesy of the surface. From the preceding discussion it should be apparent that the in-plane dependence of the scattered intensity is probably best suited for such inversions. Hence, in the following, it will be assume that the scattering measurements are performed for the in-plane configuration for one or several angles of incidence. Under this assumption, one should be able to extract the Hurst exponent from the in-plane dependence of $\left<\partial R(\pvec{q}|\pvec{k}) / \partial \Omega_s\right>/ m(\pvec{q}|\pvec{k})$ as a function of the lateral wave vector transfer $| \pvec{q}- \pvec{k} |$, in  particular, from the tail of the distribution; see Fig.~\ref{fig:analytical_tail} and 
Eq.~\eqref{eq:MDRC-diffuse-expansion}. In Fig.~\ref{fig:Hurst_extraction} we present rigorous computer simulation results~(open symbols) for the intensity scattered from self-affine Dirichlet and Neumann surfaces characterized by the Hurst exponent $H=0.70$ and two values of the topothesy $\ell=\num{e-6}\lambda$~[Figs.~\ref{fig:Hurst_extraction}(a)--(b)] and $\ell=\num{e-5}\lambda$~[Figs.~\ref{fig:Hurst_extraction}(c)--(d)]. These simulation results were obtained by the use of a Gaussian beam of width $w=10\lambda$ and the polar angles of incidence were $\theta_0=\ang{0}$~[Figs.~\ref{fig:Hurst_extraction}(a,\,c)] and $\theta_0=\ang{50}$~[Figs.~\ref{fig:Hurst_extraction}(b,\,d)]. In addition, we present two sets of results obtained on the basis of the Kirchhoff approximation.
The first set of results was generated from Eq.~\eqref{eq:MDRC-Kirchoff-Final} under the assumption of a plane incident wave and is shown as dashed black lines in Fig.~\ref{fig:Hurst_extraction}. The other set, depicted as solid orange lines in Fig.~\ref{fig:Hurst_extraction}, was calculated numerically, again within the Kirchhoff approximation, by assuming the same incident Gaussian beam as was used to generate the rigorous results presented in the same figure. It is observed from Fig.~\ref{fig:Hurst_extraction} that the Dirichlet and Neumann results are rather similar except for the largest wave vector transfers that correspond to angles of scattering that are close to grazing. Moreover, very good agreement in the central part of the scattered intensity distributions is observed between the data sets generated by the rigorous simulations and those obtained on the basis of the Kirchhoff approximation assuming the Gaussian incident beam. In particular, these results display an inverse power-law behavior of the form predicted by Eq.~\eqref{eq:MDRC-diffuse-expansion} and its use for the extraction of the Hurst exponent produce a value for the Hurst exponent in the range from  $H^\star=\num{0.70}$ to \num{0.75} depending on the region of the lateral wave vector used in the fit and if Dirichlet or Neumann data are used. For instance, from the power-law tail of the Neumann scattering data in Figs.~\ref{fig:Hurst_extraction}(a) and \ref{fig:Hurst_extraction}(c) we obtain the estimates $H^\star=\num{0.74\pm0.02}$ and $H^\star=\num{0.73\pm0.01}$, respectively. These values for $H^\star$ agree reasonably well with the value $H=0.70$ used in generate the self-affine surfaces on which the scattering calculations were based. It should be remarked that the study of Schmittbuhl \etal~\cite{Schmittbuhl1995} found that estimates of the Hurst exponent by various methods often could display errors in the range of $10\%$ and that the size of the surface analyzed  and its discretization interval significantly could affect the reliability of self-affine parameter retrieval. 

It should be pointed out that the use of a Gaussian incident beam reduce the range of $|\pvec{q}-\pvec{k}|$ values over which the scaling relation~\eqref{eq:MDRC-diffuse-expansion} holds; this is shown explicitly in Fig.~\ref{fig:Hurst_extraction} where the dashed black lines are predictions of the analytic expression in Eq.~\eqref{eq:MDRC-Kirchoff-Final} that were derived under the assumption of a plane incident wave. As was mentioned in the discussion of the results in Fig.~\ref{fig:kirchhoff_analytical_comparison} we expect plane wave illumination to be a good approximation in most practically relevant cases. Hence, by using such illumination, a larger region of $|\pvec{q}-\pvec{k}|$ values will be available for the determination of the Hurst exponent which is expected to result in more accurate estimates for this parameter.

The topothesy of the self-affine surface can also be determined from the results presented in Fig.~\ref{fig:Hurst_extraction}. In the case of a plane incident wave, the value of this parameter can be extracted from the scattered intensity in the specular direction [$\pvec{q}=\pvec{k}$], Eq.~\eqref{eq:MDRC-specular-direction}, or from the width of the specular peak, Eq.~\eqref{eq:peak-width}, given that  an estimate for the Hurst exponent already has been obtained by other means.
For instance, previously we estimated the Hurst exponent to be $H^\star=\num{0.73\pm0.01}$ from the Neumann data set in Fig.~\ref{fig:Hurst_extraction}(c), for which the use of an incident plane wave or a Gaussian beam produce almost the same specular scattered intensity. If the expression in Eq.~\eqref{eq:MDRC-specular-direction} is applied to this data set, we are led to the estimate $\ell^\star=\num{2.2E-6}\lambda$ for the topothesy of the surface when the value $H^\star$ given above was assumed. This is only $20\%$ of the value of the topothesy assumed in generating the scattering data [$\ell=\num{e-5}\lambda$]. To get reliable  estimates for the topothesy based on  Eq.~\eqref{eq:MDRC-specular-direction} unfortunately requires high precision in the estimate of the Hurst exponent. This is due to the Hurst dependent exponent that appears in Eq.~\eqref{eq:MDRC-specular-direction}. In order to increase the precision of the topothesy estimated obtained in this way, several data sets corresponding to different polar angles of incidence may be considered. In principle such an estimate can also be carried out for an incident Gaussian beam; however, in this case the scattered intensity is not available as an analytic expression that can be evaluated readily and it must instead be calculated by much more time-consuming methods. 

Instead of pursuing such an approach, it is more fruitful to perform a joint (or simultaneous) inversion of the data sets in order to reconstruct the values of the  Hurst exponent and the topothesy. To this end, the full function form of the scattering data are used. Before delving into the technical details of such an inversion procedure, we aim to investigate how well the scattering data in Fig.~\ref{fig:Hurst_extraction}, obtained by rigorous simulations and the use of a Gaussian incident beam, can be reproduced by the use of the \emph{same} incident Gaussian beam and a single scattering approach based on the Kirchhoff approximation, the same approximation used to derive the analytic result in Eq.~\eqref{eq:MDRC-Kirchoff-Final} when a plane incident wave is assumed. This is done by evaluating the equation of scattering theory within the Kirchhoff approximation~\cite{Voronovich2007} for an ensemble of surface realizations assuming the same incident Gaussian beam and self-affine parameters of the rough surfaces as were assumed in producing the rigorous results presented in Fig.~\ref{fig:Hurst_extraction}. This approach does not require solving a large system of linear equations for the field and normal derivative of the field evaluated at the surface, as is the time consuming part of the rigorous method.
In this way the data sets corresponding to the orange solid lines in Fig.~\ref{fig:Hurst_extraction} (labeled Kirchhoff) were obtained and they agree quite well with the data sets obtained by rigorous simulations for the same polar angle of incidence. There is only a noticeable discrepancy between the single and multiple scattering results for the largest values of $|\pvec{q}-\pvec{k}|$ corresponding to large polar scattering angles. The results of Fig.~\ref{fig:Hurst_extraction} testify to the accuracy of the Kirchhoff approach for the self-affine parameters that were assumed in producing the results in Fig.~\ref{fig:Hurst_extraction}.

The two Kirchhoff results presented in each of the panels in Fig.~\ref{fig:Hurst_extraction} are equivalent as they only differ in the type of incident beam that was assumed in producing them, and, as a result, in how $\left< \partial R(\pvec{q}|\pvec{k})/\partial\Omega_s\right>$ was obtained; for a plane incident beam this quantity was calculated by evaluating the analytical expression in Eq.~\eqref{eq:MDRC-Kirchoff-Final}, while for an incident  Gaussian beam, the calculation was performed on the basis of a Monte Carlo calculation assuming the Kirchhoff approximation. Based on this results, we will in the following assume that if we were able to obtain rigorous simulations results for the mean DRC using a plane incident wave, or a sufficiently wide incident beam, that the result obtained on the basis of Eq.~\eqref{eq:MDRC-Kirchoff-Final} would reproduce the corresponding rigorous results with a similar level of accuracy as was found in Fig.~\ref{fig:Hurst_extraction} when an incident Gaussian beam was used.

%
%

We are now prepared to perform \emph{nonlinear optimization} of scattering data obtained from controlled computer experiments with respect to the expression in Eq.~\eqref{eq:MDRC-Kirchoff-Final-B} in order to reconstruct the values of the Hurst exponent and the topothesy of the self-affine surface used in obtaining the scattering data. To this end, we define the \emph{cost function} to be used in the optimization
\begin{widetext}
%
\begin{align}
  \chi^2({\mathcal P})
  &=
    %
    \int\limits_{q_\parallel < \omega/c} \dfint[2]{q_\parallel}{(2\pi)^2}
    {\mathcal W}(\pvec{q}|\pvec{k})    
    \left[
    \log \left< \frac{\partial R(\pvec{q}|\pvec{k}) }{ \partial \Omega_s} \right>
    -\log \left. \left< \frac{\partial R(\pvec{q}|\pvec{k} ) }{ \partial \Omega_s} \right> \right|_{ \{H,\ell\}={\mathcal P}}
    \right]^2,
    \label{eq:cost}
\end{align}
\end{widetext}
%
where ${\mathcal P}=\{H^\star,\ell^\star\}$  is the set of parameter values that the optimization aims to determine. Here $\left< \partial R(\pvec{q}|\pvec{k}) /\partial \Omega_s \right>$ denotes the measured mean DRC while $\left< \partial R(\pvec{q}|\pvec{k} ) / \partial \Omega_s \right>|_{ \{H,\ell\}={\mathcal P}}$ is a data set obtained on the basis of Eq.~\eqref{eq:MDRC-Kirchoff-Final-B} using parameter values  $\{H,\ell\}={\mathcal P}$ and the same values for $\pvec{q}$ used in obtaining the measured data set. In writing Eq.~\eqref{eq:cost} we have defined a potential  weighting function, ${\mathcal W}(\pvec{q}|\pvec{k})$, that is different from a constant over the domain $q_\parallel < \omega/c$ only if weighted optimization is performed; we use the logarithm of the mean DRCs in the definition of the cost function in order to reduce the dynamical range that these quantities possess. In the optimization that will be performed, we have restricted ourselves to the use of  in-plane scattering data (for which $q_2=0$).

\begin{figure*}[htb!]
  \centering
  \includegraphics[width=\textwidth]{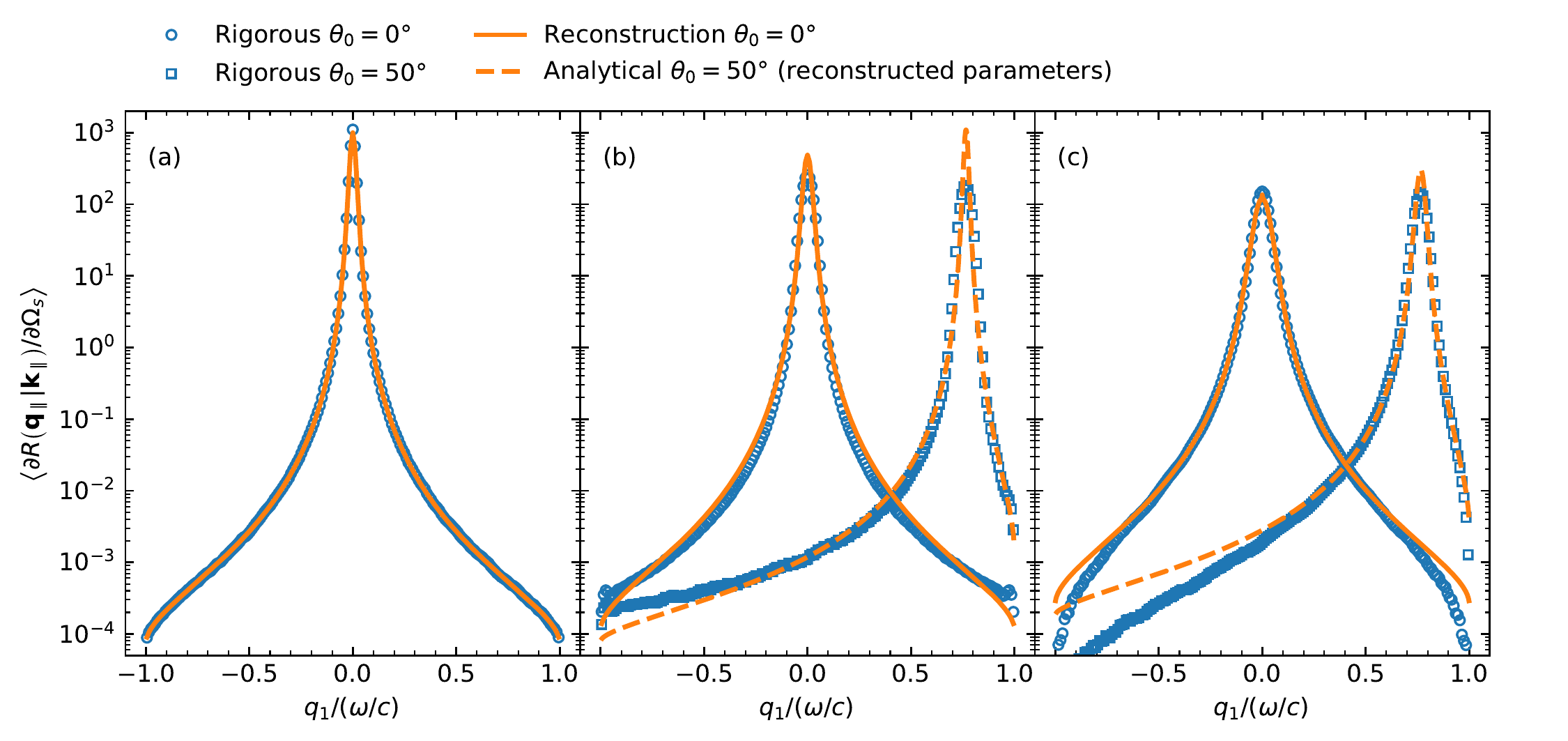} 
  \caption{
    Inversion of in-plane scattering data obtained in computer experiments for Gaussian incident beams scattered from self-affine surfaces of Hurst exponent $H=0.70$ and several values of the topothesy $\ell$.
    (a) Monte Carlo simulated input data~(blue open circles) for polar angle of incidence $\theta_0=\ang{0}$ generated within the Kirchhoff approximation for a self-affine surface of topothesy $\ell=\num{e-6}\lambda$ where $\lambda$ is the wavelength of the incident field. These data were obtained by averaging the results over $N_\zeta=\num{10000}$ surface realizations and the incident Gaussian beam had width $w=32\lambda$. The orange sold mean DRC curve was produced by inverting the input data to obtain the reconstructed self-affine parameters of values $H^\star=\num{0.70}$ and $\ell^\star=\num{1.25e-6}\lambda$.
    (b) In-plane scattering data (blue open symbols) generated by rigorous simulations for self-affine Neumann surfaces of topothesy  $\ell=\num{e-6}\lambda$ and obtained by averaging the results over $N_\zeta=\num{5000}$ surface realizations.  Incident Gaussian beams of width $w=10\lambda$ and polar angles of incidence $\theta_0=\ang{0}$ and $\theta_0=\ang{50}$ were assumed in generating these results.  Inversion of the computer generated data set for polar angle of incidence  $\theta_0=\ang{0}$ produced the orange solid line and the reconstructed self-affine parameters $H^\star=\num{0.71}$ and $\ell^\star=\num{1.41e-6}\lambda$. The dashed orange line was produced from Eq.~\eqref{eq:MDRC-Kirchoff-Final} using  the reconstructed values for $H$ and $\ell$.
(c) Same as Fig.~\protect\ref{fig:reconstruction}(b) but computer simulated data sets (open blue symbols) corresponding to a Dirichlet surface of topothesy $\ell=\num{e-5}\lambda$. The inversion of the $\theta_0=\ang{0}$ date set was performed  for $|q_1|<0.5\omega/c$; this produced the solid orange line and the reconstructed parameter values $H^\star=\num{0.71}$ and $\ell^\star=\num{7.0e-6}\lambda$. Using these values and $\theta_0=\ang{50}$ in Eq.~\eqref{eq:MDRC-Kirchoff-Final} produced the orange dashed line.
\label{fig:reconstruction}}
\end{figure*}

%
%
The first inversion that we will perform used input data obtained on the basis of the Kirchhoff approximation using a Gaussian beam of width $w=32\lambda$ and self-affine surface parameters $H=0.70$ and $\ell=\num{e-6}\lambda$; this data set is depicted by blue open symbols in Fig.~\ref{fig:reconstruction}(a). This and subsequent inversions (minimizations) started from the initial parameter values ${\mathcal P}=\{0.50,\num{e-3}\lambda\}$ and was performed using the Levenberg-Marquardt algorithm~\cite{Levenberg1944,Marquardt1963} where the elements of the Hessian that this algorithm requires were calculated by finite difference approximations. For the  weighting function we used unity for all values of its argument. 
The minimization of cost function~\eqref{eq:cost} for the input data specified above converged towards the data set given as an orange solid line in Fig.~\ref{fig:reconstruction}(a) and an excellent agreement between the input and reconstructed mean DRC curves is found. The self-affine parameter values reconstructed in this way were $H^\star=\num{0.70}$ and $\ell^\star=\num{1.25e-6}\lambda$ which are in good agreement with the values assumed when generating the input scattering data on which the inversion was based [$H=0.70$ and $\ell=\num{e-6}\lambda$]. The results presented in Fig.~\ref{fig:reconstruction}(a) demonstrate explicitly that the use of an incident Gaussian beam of width $w=32\lambda$, is sufficiently wide to produce a mean DRC that is well approximated by a mean DRC that assumes a plane incident wave and is given by Eq.~\eqref{eq:MDRC-Kirchoff-Final}. Moreover, since both the input data set and the expression in Eq.~\eqref{eq:MDRC-Kirchoff-Final} were generated on the basis of the Kirchhoff approximation, though for different types of incident beams, these results hints towards the quality that can be achieved for the reconstructed values of the self-affine parameters of the surface.

%
%
%
%

%
%
We will now turn to the more relevant and interesting case of the inversion of in-plane scattering data obtained on the basis of a rigorous and, therefore, a multiple scattering approach. For this purpose, we use the Neumann scattering data presented in Fig.~\ref{fig:Neumann_rigorous_topothesy_comparison}(a) that were obtained on the assumption of a normally incident Gaussian beam and self-affine parameters $H=0.70$ and $\ell=\num{e-6}\lambda$; this data set is presented as blue open circles in Fig.~\ref{fig:reconstruction}(b). Based on this data set, the cost function~\eqref{eq:cost} was minimized in a completely equivalent manner to what was done above to produce the orange solid line in Fig.~\ref{fig:reconstruction}(b) and the reconstructed self-affine parameter values were $H^\star=\num{0.71}$ and $\ell^\star=\num{1.41e-6}\lambda$. These values agree rather well with the self-affine parameters that characterize the surface from which the input scattering data were generated. Moreover, the angular dependence of the input and inverted mean DRC curves are also rather good; the main discrepancies between them are found around the normal scattering direction due to the different forms of incident beams that they assume and in the tails of the scattered distributions (probably caused by multiple scattering). In addition to the two mean DRC curves in Fig.~\ref{fig:reconstruction}(b) for $\theta_0=\ang{0}$, we also in this figure present, for reasons of comparison, the mean DRCs for the polar angle of incidence $\theta_0=\ang{50}$. Here the blue open square symbols refer to result obtained for the same surface by rigorous computer simulations for this polar angle of incidence while the dashed orange line was produced from Eq.~\eqref{eq:MDRC-Kirchoff-Final} assuming the self-affine parameters previously obtained by reconstruction of the data set that corresponds to normal incidence. 

%
%
The last inversion example that we will give, is based on the Dirichlet scattering data from Fig.~\ref{fig:Dirichlet_rigorous_topothesy_comparison}(b) for normal incidence~[blue open circles in Fig.~\ref{fig:reconstruction}(c)]. These data were generated by rigorous simulations in an equivalent manner to how the input data from our previous examples were obtained; the main difference is that now the topothesy is $\ell=\num{e-5}\lambda$. In order to invert this data set, we used a weighting function in Eq.~\eqref{eq:cost} that rapidly tapered off outside $|q_1| < 0.5 \omega/c$. In this way the mean DRC shown as an orange solid line in  Fig.~\ref{fig:reconstruction}(c) was obtained and it corresponds to the reconstructed parameters  $H^\star=\num{0.71}$ and the topothesy $\ell^\star=\num{7.0e-6}\lambda$. Also in this case, we have for reasons of comparison added results for $\theta_0=\ang{50}$, and the orange dashed line is generated from Eq.~\eqref{eq:MDRC-Kirchoff-Final} assuming the self-affine parameters reconstructed from the data set for normal incidence. The results in Fig.~\ref{fig:reconstruction}(c) show reasonable agreement between the simulated and analytic results, and the reconstructed self-affine parameters are in good  agreement with the parameters assumed in producing the simulation data.

\begin{figure}[tb!]
  \centering
  \includegraphics[width=\columnwidth]{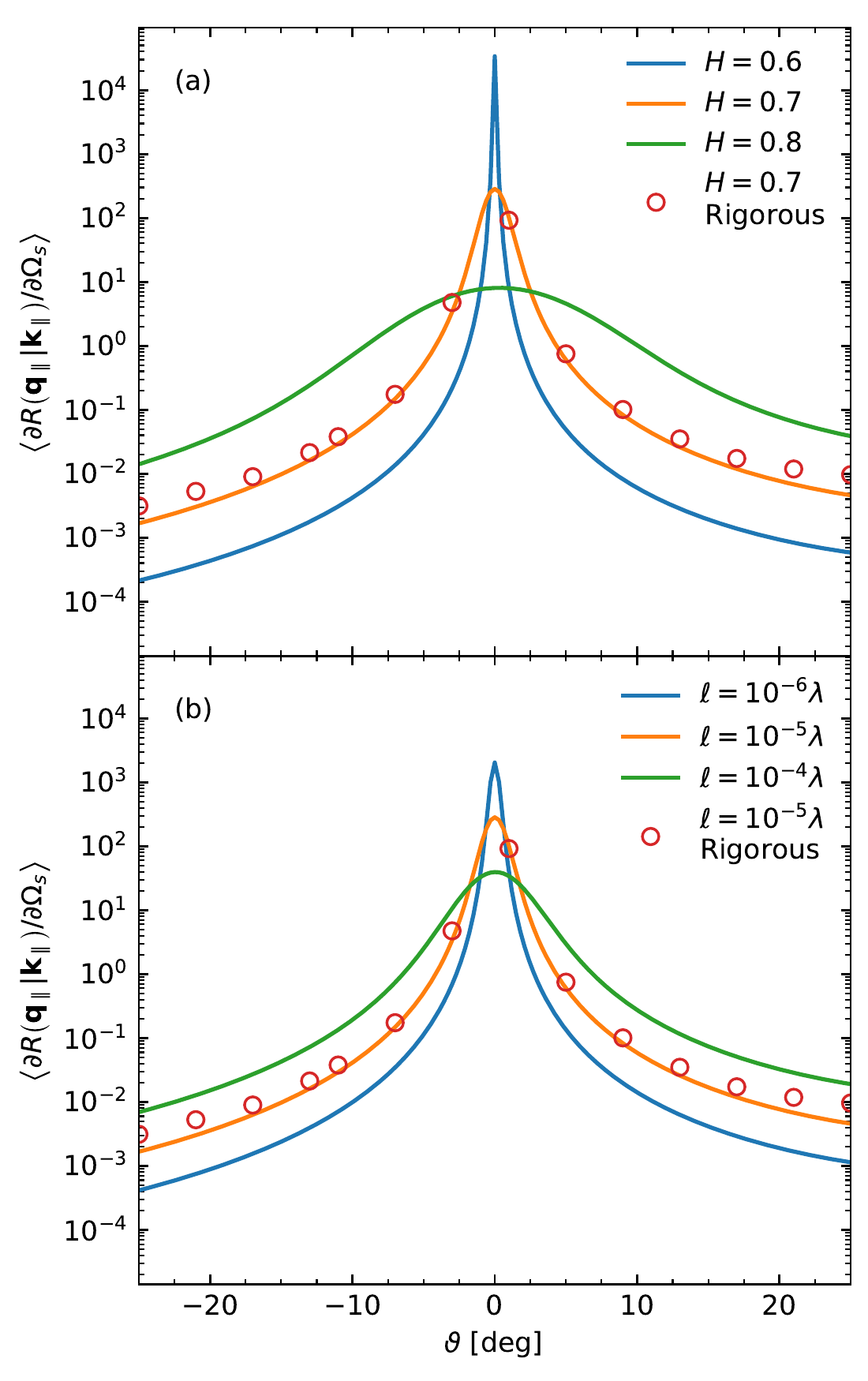}
  \caption{Rocking scan curves for self-affine surfaces characterized by (a) several values of the Hurst exponent (see legend) and topothesy $\ell=\num{e-5}\lambda$; and (b) Hurst exponent $H=0.70$ and several values of the topothesy $\ell$ (see legend) for  $\theta_0=\theta_s=\ang{45}$.
  The solid lines were obtained on the basis of Eq.~\eqref{eq:MDRC-Kirchoff-Final} while the red open symbols were produced by rigorous computer simulations assuming a self-affine Neumann surface of parameters $H=0.70$ and $\ell=\num{e-5}\lambda$ and an incident Gaussian beam of width $w=10\lambda$.
  The rocking angle $\vartheta$ is defined so that the effective polar angle of incidence is $\theta_0+\vartheta$, and the effective polar angle of scattering is $\theta_s-\vartheta$.}
  \label{fig:rocking_semilog} 
\end{figure}

%
%
\medskip
Up till now we have discussed methods for the  extraction of self-affine parameters that required angular resolved scattering intensity measurements for one or several angles of incidence, or specular intensity measurements for many angles of incidence. Without access to automated specialized equipment, such measurements are time-consuming to perform due to the alignment procedure that they require. An alternative experimental configuration will now be described that assumes fixed source and detector positions and measurements based on this configuration should therefore be both simpler and require less sophisticated (and therefore cheaper) equipment  to perform. This configuration, here referred to as the ``rocking scan'' geometry, consists of given polar angles of incidence and scattering relative to the laboratory frame, $\theta_0$ and $\theta_s$, respectively, and a tilt of the sample around a normal vector to the plane of incidence that lies in the mean plane of the surface. Such a tilt of the sample through an angle $\vartheta$ in the counter-clockwise direction, realizable by the use of a  rotation stage, for instance, will effectively change the (in-plane) polar angle of incidence from $\theta_0$ to $\theta_0'=\theta_0+\vartheta$, and at the same time change the (in-plane) polar angle of scattering from $\theta_s$ to $\theta_s'=\theta_s-\vartheta$. Equivalently, these polar angles correspond to the lateral wave vectors of incidence and scattering, $\pvec{k}'$ and $\pvec{q}'$, respectively,  defined  by  Eq.~\eqref{eq:kq-parallel} after $\theta_0$ and $\theta_s$ have been replaced by their primed equivalents and the relations $\phi_0'=\phi_0$ and  $\phi_s'=\phi_s$ have been used. The lateral wave vector transfer that corresponds to a given rocking (or tilt) angle $\vartheta$ thus becomes
\begin{align}
  \pvec{Q}'(\vartheta)
   \equiv
  \pvec{q}' - \pvec{k}' 
  &=   
    \pvec{Q}\cos\vartheta
    -
    \pvecUnit{k}
    \left[
    \alpha_0(q_\parallel) + \alpha_0(k_\parallel)
    \right]
    \sin\vartheta,
    \label{eq:Q_vartheta}
\end{align}
where $\pvec{Q}=\pvec{q}-\pvec{k}$. Figure~\ref{fig:rocking_semilog}(a) presents the in-plane dependence of the mean DRC as function of the rocking angle $\vartheta$ when $\theta_0=\theta_s=\ang{45}$, or equivalently $\pvec{Q}=\vec{0}$. The solid lines in this figure were obtained on the basis of the analytic expression in Eq.~\eqref{eq:MDRC-Kirchoff-Final} and under the assumption of a self-affine surface characterized by topothesy $\ell=\num{E-5}\lambda$ and a Hurst exponent $H$ as indicated in the legend. The red symbols in Fig.~\ref{fig:rocking_semilog}(a) were obtained by rigorous computer simulations of the scattering from Neumann surface of parameters $H=0.70$ and $\ell=\num{e-5}\lambda$ when the incident beam was assumed to have width $w=10\lambda$.
Similar results are presented in Fig.~\ref{fig:rocking_semilog}(b) but here the value of the  Hurst exponent  is fixed to $H=0.70$ and several values of the topothesy are considered as specified in the legend. Note that the symbols and the orange lines in Figs.~\ref{fig:rocking_semilog}(a) and \ref{fig:rocking_semilog}(b) depict the same data.  The results presented in Fig.~\ref{fig:rocking_semilog} demonstrate that the rocking scan curves display dependencies with respect to both the Hurst exponent and the topothesy of the self-affine surface that are suitable for performing reconstruction of these parameters based on measured rocking scan curves. Even if we will not do such reconstruction here, it is noted that the data obtained by rigorous simulations, red symbols in Fig.~\ref{fig:rocking_semilog}, agree rather favorably with what is obtained from the analytic expression~\eqref{eq:MDRC-Kirchoff-Final}.

%
%
%

%
%
\medskip
So far we have discussed several of the many methods that exist for extracting the values of the  parameters that characterize a self-affine surface from the (in-plane) scattered intensity that it gives rise to when an incident beam impinges on it. Before closing this study, we will mention an additional method for doing so that is based on \emph{rescaling}. This is particularly interesting since scaling is at the hearth of the self-affine property. Moreover, a similar property obtained for the scattered intensity from  one-dimensional self-affine surfaces has previously been established~\cite{Simonsen1999-1,Simonsen1999-3}, so for reasons of completeness, we want here to establish a similar scaling relation of the intensity that is scattered from two-dimensional self-affine surfaces. Below we will see that if the in-plane scattering data are rescaled in a certain manner, they should, if single scattering is dominating, collapse onto a \emph{master curve}  that only depends on the Hurst exponent of the self-affine surface and not on, for instance, the topothesy or the polar angle of incidence used when performing the measurements. To see how this comes about, we return to Eq.~\eqref{eq:MDRC-Kirchoff-Final-B} that we for convenience rewrite in the alternative form
\begin{subequations}
  \label{eq:MDRC-factorized}
\begin{align}
  \left< \frac{\partial R(\pvec{q}|\pvec{k}) }{ \partial \Omega_s} \right>   
  &=
    M( \pvec{q}, \pvec{k}; H, \ell)\,
      {\cal L}_{2H}\!\! 
      \left(  
    \chi( \pvec{q}, \pvec{k}; H, \ell)
        ; 
        \frac{1}{2}
    \right),
    \label{eq:MDRC-factorized-A}
\end{align}
where we have defined the functions
\begin{align}
  \label{eq:MDRC-factorized-B}
  M( \pvec{q},& \pvec{k}; H, \ell )
                \nonumber \\
  &=
   \frac{
     \left(\omega/c\right)
    \left[ \left(\omega/c\right)^2 + \alpha_0(q_\parallel) \alpha_0(k_\parallel) - \pvec{q} \cdot \pvec{k} \right]^2
    }{
    \alpha_0(k_\parallel) \left[ \alpha_0(q_\parallel) + \alpha_0(k_\parallel)  \right]^{(2+2H)/H} \ell^{(2-2H)/H}
    }
\end{align}
and
\begin{align}
  \label{eq:MDRC-factorized-C}
  \chi( \pvec{q}, \pvec{k}; H, \ell )
  &=
    \frac{ 
    \left| \pvec{q}-\pvec{k} \right| 
    }{ 
    \left[ \alpha_0(q_\parallel) + \alpha_0(k_\parallel)   \right]^{1/H}    
    \ell^{(1-H)/H}     
    }.
\end{align}
\end{subequations}
From the expressions in Eq.~\eqref{eq:MDRC-factorized}, it should be apparent that if we plot $\left<\partial R(\pvec{q}| \pvec{k}) / \partial \Omega_s \right>   M^{-1}( \pvec{q}, \pvec{k}; H, \ell )$ against $\chi( \pvec{q}, \pvec{k}; H, \ell )$, the result for a given Hurst exponent, $H$, should collapse onto the \textit{master curve} ${\mathcal L}_{2H}(\chi; 1/2)$ for \textit{any} polar angle of incidence $\theta_0$ (or $\pvec{k}$) and any value of the topothesy $\ell$; at least, this is the case within the single scattering approximation. If a double logarithmic scale is used for making the  plot, the Hurst exponent can be extracted from the slope of the power-law tail of the L\'evy distribution ${\mathcal L}_{2H}(\cdot)$ onto which the scattering data scaled in this manner should collapse. According to Eq.~\eqref{app:eq:asymptotic-expansion-2}, or Eq.~\eqref{eq:MDRC-diffuse-expansion}, the slope of the resulting power-law tail should be $-2-2H$ from which the Hurst exponent of the surface can be obtained. The data collapse relies on the expression in Eq.~\eqref{eq:MDRC-Kirchoff-Final} being an accurate description of the mean DRC for the surface, or, in other words, that the scattering is dominated by single-scattering. This may, or may not, be the case depending on the values of the self-affine parameters of the surface that are assumed.

%
%
\begin{figure}[tb!]
  \centering
  \includegraphics[width=\columnwidth]{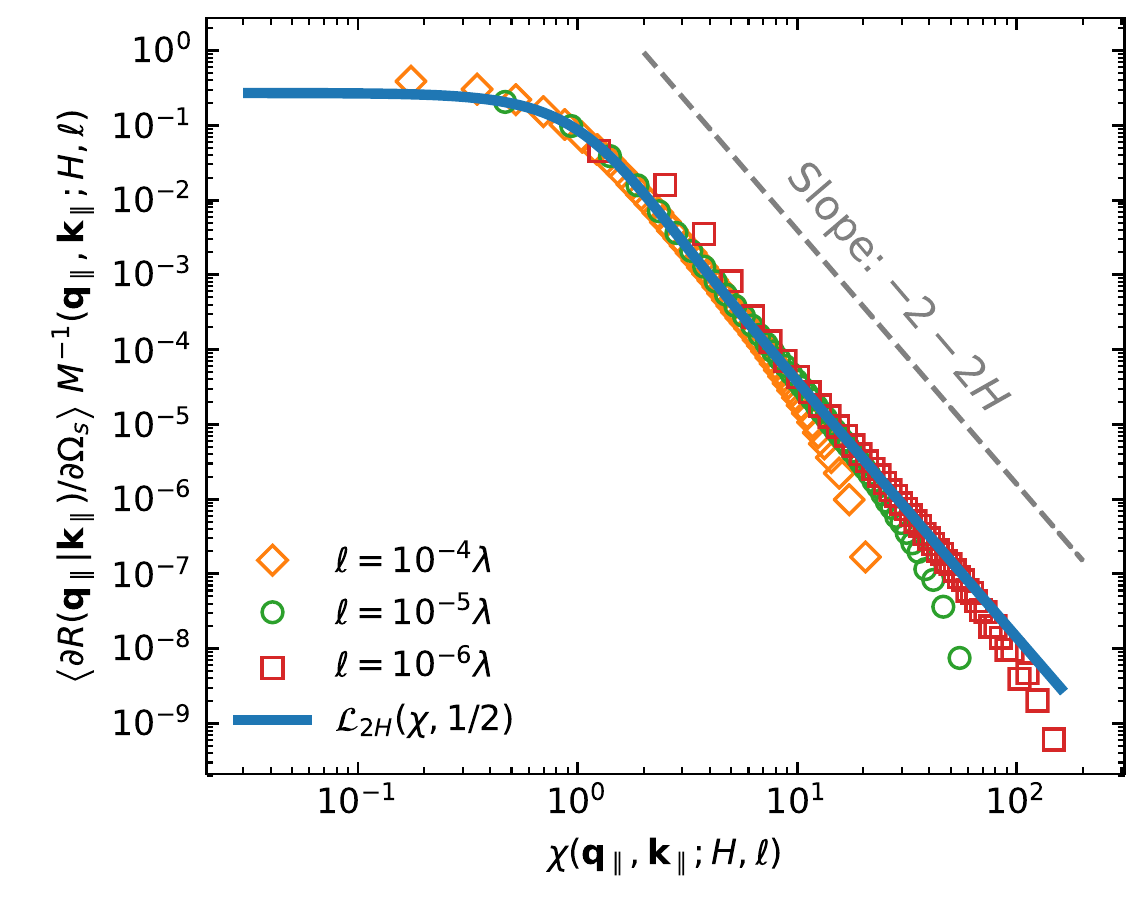}
  \caption{\label{fig:data_collapse}
    Rescaled in-plane mean DRCs from Fig.~\ref{fig:Dirichlet_rigorous_topothesy_comparison} that were generated by rigorous computer simulations of $w=10\lambda$ wide normally incident Gaussian beams scattered from self-affine Dirichlet surfaces of Hurst exponent $H=0.70$ and topothesies $\ell$ as indicated in the legend (open symbols). The rescaling was done according to Eq.~\eqref{eq:MDRC-factorized} so what is presented is $\left<\partial R(\pvec{q}| \pvec{k}) / \partial \Omega_s \right>   M^{-1}( \pvec{q}, \pvec{k}; H, \ell )$ as function of $\chi( \pvec{q}, \pvec{k}; H, \ell )$. The blue solid line is the ``master curve'' ${\mathcal L}_{2H}(\chi,1/2)$ that is predicted by Eq.~\eqref{eq:MDRC-factorized}.
    The data collapse brings out the L\'evy shape in accordance with Eq.~\eqref{eq:MDRC-Kirchoff-Final}.
  }
\end{figure}
%
%

Figure~\ref{fig:data_collapse} presents as open symbols the rescaled quantity $\left<\partial R(\pvec{q}| \pvec{k}) / \partial \Omega_s \right>   M^{-1}( \pvec{q}, \pvec{k}; H, \ell )$ as function of $\chi( \pvec{q}, \pvec{k}; H, \ell )$ obtained on the basis of the mean DRC data sets in Fig.~\ref{fig:Dirichlet_rigorous_topothesy_comparison} for normal incidence that were generated by rigorous computer simulations of a $w=10\lambda$ wide incident Gaussian beam scattered from a self-affine Dirichlet surface of Hurst exponent $H=0.70$. In Fig.~\ref{fig:data_collapse} the  solid line represents the ``master curve'' ${\mathcal L}_{2H}(\chi,1/2)$ that is predicted from Eq.~\eqref{eq:MDRC-factorized} for a plane incident wave. Note that we could in producing Fig.~\ref{fig:data_collapse} also have included data for $\theta_0=\ang{50}$ and we have found that also these  would  end up on the master curve but these results are not shown here.

From the results presented in Fig.~\ref{fig:data_collapse},  we find that the idea of a data collapse of the rescaled mean DRC data onto a master curve works rather well for a wide range of topothesies and angles of incidence even when the incident beam is not a plane wave. Still some deviations from this L\'evy master curve is noticeable and expected, in particular, for the larges values of the topothesy and/or for grazing angles of incidence and scattering which corresponds to situations when multiple scattering effects are expected to become important. The results in Fig.~\ref{fig:data_collapse}, obtained for the scattering from two-dimensional self-affine surfaces, should be compared to similar results established for the scattering from one-dimensional self-affine surfaces; for instance, see Figs.~3 and 7 in Refs.~\onlinecite{Simonsen1999-1} and \onlinecite{Simonsen1999-3}, respectively.

\section{Conclusions}
\label{Sec:Conculusions}
Within the Kirchhoff approximation, we have derived an analytic expression for the mean differential reflection coefficient for the wave scattering from two-dimensional self-affine Dirichlet and Neumann surfaces. Under the assumption of a plane incident wave, the angular dependence of this quantity is expressed in terms of the isotropic, bivariate ($\alpha$-stable) L\'evy distribution of stability parameter $2H$ where $H$ is the Hurst exponent of the surface. We find good agreement between the prediction of the analytic expression that we derived for the mean differential reflection coefficient and the results for the scattered intensity obtained from rigorous computer simulations. Motivated by the analytic results, we present several methods for reconstructing from the scattered intensity the values of the parameters that characterize the self-affine surface. We hope that the results of this study may motivate experimentalists to apply inverse scattering techniques more extensively for the large-scale characterizations of self-affine surface morphologies.

\begin{acknowledgments}
  The work of T.N. received support from the Research Council of Norway, Fripro Project No.~213453. The research of I.S. was supported in part by the Research Council of Norway~(Contract 216699) and the French National Research Agency (ANR-15-CHIN-0003). This research was supported in part by NTNU and the Norwegian metacenter for High Performance Computing (NOTUR) by the allocation of computer time.
\end{acknowledgments}

\appendix
\section{Properties of the isotropic bivariate L\'evy distribution}
\label{App:Levy}

Here some useful relations for the isotropic bivariate L\'evy distribution, ${\cal L}_\alpha ( \pvec{Q}; \gamma )$  will be derived. This distribution is defined by Eq.~\eqref{eq:Levy} as the two-dimensional Fourier transform of its characteristic function --- the \textit{stretched exponential} $\exp(-\gamma|\pvec{v}|^\alpha)$.
Since this latter function is circular symmetric, it follows that the isotropic bivariate L\'evy distribution may be expresses in the alternative form
 \begin{align}
   \label{eq:Levy-Hankel}
  {\cal L}_\alpha(\pvec{Q}; \gamma) 
  &= 
  \frac{1}{2\pi} \int_0^\infty \dint{v_\parallel} v_\parallel \operatorname{J}_0\left( Q_\parallel v_\parallel \right) 
  \exp\left(- \gamma v_\parallel^\alpha \right)
    \nonumber \\
   &
   \equiv
   {\cal H}_0\left[\frac{ \exp( - \gamma v_\parallel^\alpha) }{ 2\pi } \right](Q_\parallel).
\end{align}
Here  $\operatorname{J}_0(\cdot)$ denotes the Bessel function of the first kind and order zero and ${\cal H}_0[f]$ represents the zero-order Hankel (or Fourier-Bessel)  transform~\cite{Secada1999} of a function $f(v_\parallel)$ which is defined implicitly by Eq.~\eqref{eq:Levy-Hankel}. The fact that the \textit{isotropic} bivariate  L\'evy distribution can be expressed as a zero-order Hankel transform of the characteristic function, is a direct consequence of a general property of the two-dimensional Fourier transform of a circular symmetric function~\cite{Book:Jerri1992}. It shows explicitly that the distribution ${\cal L}_\alpha(\pvec{Q}; \gamma)$ depends on its argument $\pvec{Q}$ only through its norm $Q_\parallel=|\pvec{Q}|$ but not its direction, as is to be expected for an isotropic distribution. When ${\cal L}_\alpha(\pvec{Q}; \gamma) $ is evaluated numerically, the form~\eqref{eq:Levy-Hankel} is preferred over the form~\eqref{eq:Levy}, since the former expression can be evaluated more efficiently numerically than the expression that appears in the latter due to the one-dimensional integral that the former contains compared to the two-dimensional integral that is present in the latter form. 

L\'evy distributions of given stability parameter $\alpha$ but different scale parameters $\gamma$ are related. By making the change of variable $\pvec{v}=\pvec{u}/\bar{\gamma}^{1/\alpha}$ (with $\bar{\gamma}>0$) in Eq.~\eqref{eq:Levy} or Eq.~\eqref{eq:Levy-Hankel}, the following \textit{scaling relation} is obtained
\begin{align}
  \label{eq:Levy-scaling}
  {\cal L}_\alpha\left( \pvec{Q}; \gamma \right) 
  &= 
  \frac{1}{\bar{\gamma}^{2/\alpha}} \; 
  {\cal L}_\alpha\left( \frac{\pvec{Q}}{\bar{\gamma}^{1/\alpha}}; \frac{\gamma}{\bar{\gamma}} \right). 
\end{align}
This relation is useful for modifying the scale parameter of the isotropic bivariate L\'evy distribution.

\medskip
For a general stability parameter $\alpha$, the L\'evy distribution ${\cal L}_\alpha(\pvec{Q}; \gamma)$ has several important series expansions. For instance, for arguments of small norm $Q_\parallel=|\pvec{Q}|$, expanding the Bessel function that appears in Eq.~\eqref{eq:Levy-Hankel} around zero and integrating the resulting series term-by-term~\cite{Book:Abramowitz1964,NIST:DLMF}, leads to
 \begin{align}
  {\cal L}_\alpha&\left( \pvec{Q}; \gamma \right)
  \nonumber \\                 
  &=  \frac{1}{2\pi\alpha\gamma^{2/\alpha}}
     \sum_{m=0}^\infty \frac{(-1)^m}{2^{2m}(m!)^2} \Gamma\left(\frac{2m+2}{\alpha}\right) 
     \left( \frac{Q_\parallel}{\gamma^{1/\alpha}}\right)^{2m}
  \nonumber \\
  &\approx
  \frac{1}{2\pi\alpha\gamma^{2/\alpha}}
  \Bigg[   
                                             \Gamma \left( \frac{2}{\alpha} \right)  
    - \frac{1}{4}   \Gamma \left( \frac{4}{\alpha} \right)  
                    \left( \frac{Q_\parallel}{\gamma^{1/\alpha}}  \right)^2
    \nonumber \\
    & \qquad \qquad \qquad 
    + \frac{1}{64}  \Gamma \left( \frac{6}{\alpha} \right) 
                    \left( \frac{Q_\parallel}{\gamma^{1/\alpha}}  \right)^4
    - \ldots
    \Bigg];
      \nonumber \\ & \qquad \qquad \qquad  \qquad
                     \qquad \qquad 
    \qquad \frac{Q_\parallel}{\gamma^{1/\alpha}} \ll 1,
    \label{app:eq:small-argument-expansion}
\end{align}
where $\Gamma(\cdot)$ denotes the Gamma function~\cite[Ch.~5]{NIST:DLMF}.

On the other hand,  for $0<\alpha<2$, a large argument asymptotic expansion of ${\cal L}_\alpha(\pvec{Q}; \gamma)$ displays the inverse power-law tail  $Q_\parallel^{-2-\alpha}$. This can be established from Eq.~\eqref{eq:Levy-Hankel} by first expanding into a power series in $v_\parallel$ the stretched exponential function $\exp(-\gamma|v_\parallel|^\alpha)$ that is present in the integrand, reversing the order of the integration and summation, and formally performing term-by-term integration over $v_\parallel$ of the resulting expression to produce~\cite[Integral 6.561-14]{Book:Gradshteyn2007}
\begin{align}
  \label{app:eq:asymptotic-expansion-1}
  {\cal L}_\alpha(\pvec{Q}; \gamma) 
  &=
  \frac{1}{2\pi} 
  \sum_{m=0}^\infty \frac{ (-\gamma)^m }{m!} 
  \frac{ 2^{1+\alpha m} }{ Q_\parallel^{2+\alpha m} }
  \frac{ \Gamma\left( 1+\frac{\alpha m}{2} \right) }{ \Gamma\left( -\frac{\alpha m}{2} \right)  }.
\end{align}
Since the function $\Gamma(z)$ tends to infinity for vanishing argument~\cite[Ch.~5]{NIST:DLMF}, 
the $m=0$ term will not contribute to the sum in Eq.~\eqref{app:eq:asymptotic-expansion-1}. Hence, the leading term in the large argument asymptotic expansion of ${\cal L}_\alpha(\pvec{Q}; \gamma)$ becomes
\begin{align}
  \label{app:eq:asymptotic-expansion-2}
  {\cal L}_\alpha(\pvec{Q}; \gamma)
  &
  \sim
  - 
  \frac{ 2^\alpha }{ \pi }
  \frac{ \Gamma\left( 1+\frac{\alpha}{2} \right) }{ \Gamma\left( -\frac{\alpha}{2} \right)  }
    \frac{ \gamma }{ Q_\parallel^{2+\alpha} }
     ,
  \qquad
  \frac{Q_\parallel}{\gamma^{1/\alpha}} \gg 1, 
  %
  %
  \nonumber \\ &  
  =
  - 
  \frac{ 2^\alpha }{ \pi \gamma^{2/\alpha} }
  \frac{ \Gamma\left( 1+\frac{\alpha}{2} \right) }{ \Gamma\left( -\frac{\alpha}{2} \right)  }
  \left(
  \frac{Q_\parallel}{ \gamma^{1/\alpha} }
  \right)^{-2-\alpha}
  .               
\end{align}
The function $\Gamma(-\alpha/2)$ is strictly negative for $0<\alpha<2$, so the right-hand-side of Eq.~\eqref{app:eq:asymptotic-expansion-2} is guaranteed to be non-negative, as it has to be for a probability distribution function.  However, we prefer to rewrite expression~\eqref{app:eq:asymptotic-expansion-2} into an alternative form where the negative sign does not appear. To this end, we take advantage of Euler's reflection formula for the $\Gamma$-function~\cite[Formula 5.5.3]{NIST:DLMF}  
$\Gamma(z)\Gamma(1-z) = \pi/ \sin(\pi z)$ (for $z\neq 0,\pm 1,\ldots$). 
Using this relations with $z=-\alpha/2$ and the fact that the sinusoidal is an antisymmetric function of its argument, enables us to rewrite the expression in  Eq.~\eqref{app:eq:asymptotic-expansion-2} in the form [$0<\alpha<2$]        
\begin{align}
  \label{eq:Levy-large-agument-expansion}
    %
  {\cal L}_\alpha( \pvec{Q}; \gamma) 
  &\sim
    \frac{
    2^\alpha \Gamma^2\!\left( 1+\frac{\alpha}{2} \right)
    }{ 
    \pi^2 \, \gamma^{2/\alpha}
    }
    \left(
    \frac{Q_\parallel}{ \gamma^{1/\alpha} }
    \right)^{-2-\alpha}
    \sin\left( \frac{\pi\alpha}{2} \right),
    \nonumber \\ & \qquad \qquad \qquad \qquad \qquad
    \qquad \frac{Q_\parallel}{\gamma^{1/\alpha}} \gg 1
     \nonumber \\
   &=
      2^\alpha \gamma
      \left(
      \frac{
      \Gamma\!\left( 1+\frac{\alpha}{2} \right)
      }{ 
      \pi
      }
      \right)^2
      \frac{
      \sin\left( \frac{\pi\alpha}{2} \right)
      }{
      Q_\parallel^{2+\alpha} 
     }
     .
\end{align}
Equation~\eqref{eq:Levy-large-agument-expansion} represents the generalization to isotropic bivariate L\'evy distributions of the more well-known large argument asymptotic expansion of the univariate symmetric L\'evy distribution, known as the   ``Wintner's expansion''~\cite{Wintner1941,Book:Gnedenko1954,Montroll1984,Wuttke2012}. 
It should be noted that for the case $\alpha=2$, the pdf ${\cal L}_2(\cdot)$ equals the bivariate Gaussian distribution and thus decays exponentially with increasing $Q_\parallel^2$.

%
\bibliographystyle{aipnum4-1}
\bibliography{$HOME/Archive/Papers/BIBLIOGRAPHY,$HOME/Archive/Papers/BOOKS,$HOME/Adm/CV/PubList/Simonsen-Publications,paper2014-04} 
%

\end{document}